\documentclass[reprint,superscriptaddress,
 amsmath,amssymb,
 aps,
pra,
]{revtex4-2}

\usepackage{graphicx}% Include figure files
\usepackage{dcolumn}% Align table columns on decimal point
\usepackage{bm}% bold math
\usepackage{physics}
\usepackage{amsfonts, amsmath}
\usepackage{dsfont}
\usepackage{amsthm}
\usepackage{bm}
\usepackage{bbold}
\usepackage{hyperref}% add hypertext capabilities
\usepackage{color}
\usepackage{lipsum,babel}
\usepackage[normalem]{ulem}
\DeclareMathOperator*{\E}{\mathbb{E}}

\DeclareMathOperator*{\Prob}{Pr}

\newtheorem{theorem}{Theorem}
\newtheorem{corollary}{Corollary}

\newtheorem{lemma}{Lemma}

\begin{document}
\definecolor{navy}{RGB}{46,72,102}
\definecolor{pink}{RGB}{219,48,122}
\definecolor{grey}{RGB}{184,184,184}
\definecolor{yellow}{RGB}{255,192,0}
\definecolor{grey1}{RGB}{217,217,217}
\definecolor{grey2}{RGB}{166,166,166}
\definecolor{grey3}{RGB}{89,89,89}
\definecolor{red}{RGB}{255,0,0}

\preprint{APS/123-QED}

\title{Quantum Metrological Power of Continuous-Variable Quantum Networks}
\author{Hyukgun Kwon}
\affiliation{Department of Physics and Astronomy, Seoul National University, Seoul 08826, Republic of Korea}
\author{Youngrong Lim}
\affiliation{School of Computational Sciences, Korea Institute for Advanced Study, Seoul 02455, Korea}
\author{Liang Jiang}
\affiliation{Pritzker School of Molecular Engineering, The University of Chicago, Chicago, Illinois 60637, USA}
\author{Hyunseok Jeong}
\email{h.jeong37@gmail.com}
\affiliation{Department of Physics and Astronomy, Seoul National University, Seoul 08826, Republic of Korea}
\author{Changhun Oh}
\email{changhun@uchicago.edu}
\affiliation{Pritzker School of Molecular Engineering, The University of Chicago, Chicago, Illinois 60637, USA}

\begin{abstract}
We investigate the quantum metrological power of typical continuous-variable (CV) quantum networks. 
Particularly, we show that most CV quantum networks provide an entanglement to quantum states in distant nodes that enables one to achieve the Heisenberg scaling in the number of modes for distributed quantum displacement sensing, which cannot be attained using an unentangled probe state.
Notably, our scheme only requires local operations and measurements after generating an entangled probe using the quantum network.
In addition, we find a tolerable photon-loss rate that maintains the quantum enhancement.
Finally, we numerically demonstrate that even when CV quantum networks are composed of local beam splitters, the quantum enhancement can be attained when the depth is sufficiently large.
\end{abstract}
              
\maketitle

Quantum metrology is a study on advantages of quantum resources for parameter estimation~\cite{giovannetti2001quantum, giovannetti2004quantum, giovannetti2006quantum, Giovannetti2011, braun2018quantum, Pirandola2018}.
In many years, nonclassical features of quantum probes have been shown to achieve a better sensitivity than any classical means.
Especially in continuous-variable (CV) systems, a squeezed state, one of the most representative nonclassical states, elevates the sensitivity of optical interferometers~\cite{Demkowicz-Dobrzanski2015, oh2019optimal0} including gravitational wave detectors~\cite{caves1981quantum, abadie2011gravitational, aasi2013enhanced}.
In addition, enhanced phase estimation using a squeezed state has been implemented in many experiments~\cite{yonezawa2012quantum, berni2015ab, yu2020quantum}. 

More recently, besides quantum enhancement from a local system, much attention has been paid to employ a metrological advantage from entanglement between distant sites.
Particularly, distributed quantum sensing has been proposed and extensively studied to enhance the sensitivity by exploiting quantum entanglement constituted by a quantum network for estimating parameters in distant nodes~\cite{komar2014quantum, proctor2018multiparameter, ge2018distributed, zhuang2018distributed, guo2020distributed, xia2020demonstration, oh2020optimal, zhao2021field, PhysRevX.9.041023, PhysRevX.11.021047,oh2021distributed}.
For example, a single-mode squeezed vacuum state distributed by a balanced beam splitter network (BSN) was shown to enable estimating the quadrature displacement with a precision up to the Heisenberg scaling in the number of modes, which cannot be achieved without entanglement \cite{zhuang2018distributed}.
Such an enhancement has also been found in distributed quantum phase sensing~\cite{ge2018distributed, guo2020distributed, oh2020optimal,oh2021distributed}.
Furthermore, the enhancement from entanglement between nodes has been experimentally demonstrated in various tasks~\cite{guo2020distributed, xia2020demonstration, Liu2021, zhao2021field}. 

While particular CV quantum networks provide an enhancement for distributed sensing, it is unclear whether general quantum networks are beneficial. 
Since quantum entanglement between distant nodes is the key to improving the sensitivity in many cases, investigating what kinds of quantum networks are advantageous for distributed sensing is crucial fundamentally and practically.
To answer similar questions such as the usefulness of general quantum states, Ref.~\cite{oszmaniec2016random} has initiated a study for quantum enhancement from typical quantum states by considering the role of interparticle entanglement for quantum phase estimation and shown advantages of typical bosonic random states for quantum phase estimation.
%\ccor{(In the literature, there has been a study for quantum metrological enhancement of typical random quantum states and shown advantages of bosonic random states for quantum phase estimation~\cite{oszmaniec2016random}.)}

%\cor{Motivated by Ref.~\cite{oszmaniec2016random} which studies quantum metrological usefulness of typical random states in the phase sensing via random unitary,}
In this Letter, motivated by Ref.~\cite{oszmaniec2016random},
we study global random CV networks and show that typical CV quantum networks provide quantum metrological enhancement. 
More specifically, we prove that most CV quantum networks except for an exponentially small fraction in the number of modes enable us to achieve the Heisenberg scaling in the number of modes for a distributed quantum displacement sensing scheme.
Since we focus on the Heisenberg scaling in the number of sensor nodes, the intermode entanglement is the key resource.
On the other hand, Ref.~\cite{oszmaniec2016random} investigates the Heisenberg scaling in the number of particles for quantum phase estimation with interparticle entanglement.
In addition, we show that local operations after an input quantum state undergoes a CV quantum network are essential for the enhancement because the Heisenberg scaling cannot be attained without them with a high probability.
We then study the effect of photon loss and find tolerable loss amount that maintains the Heisenberg scaling.
Furthermore, we numerically demonstrate that quantum networks composed of local-random beam splitters also render the Heisenberg scaling for distributed displacement sensing on average within a depth proportional to $M^2$ with $M$ being the number of modes.

\begin{figure}[t]
    \centering
    \includegraphics[width=0.95\linewidth]{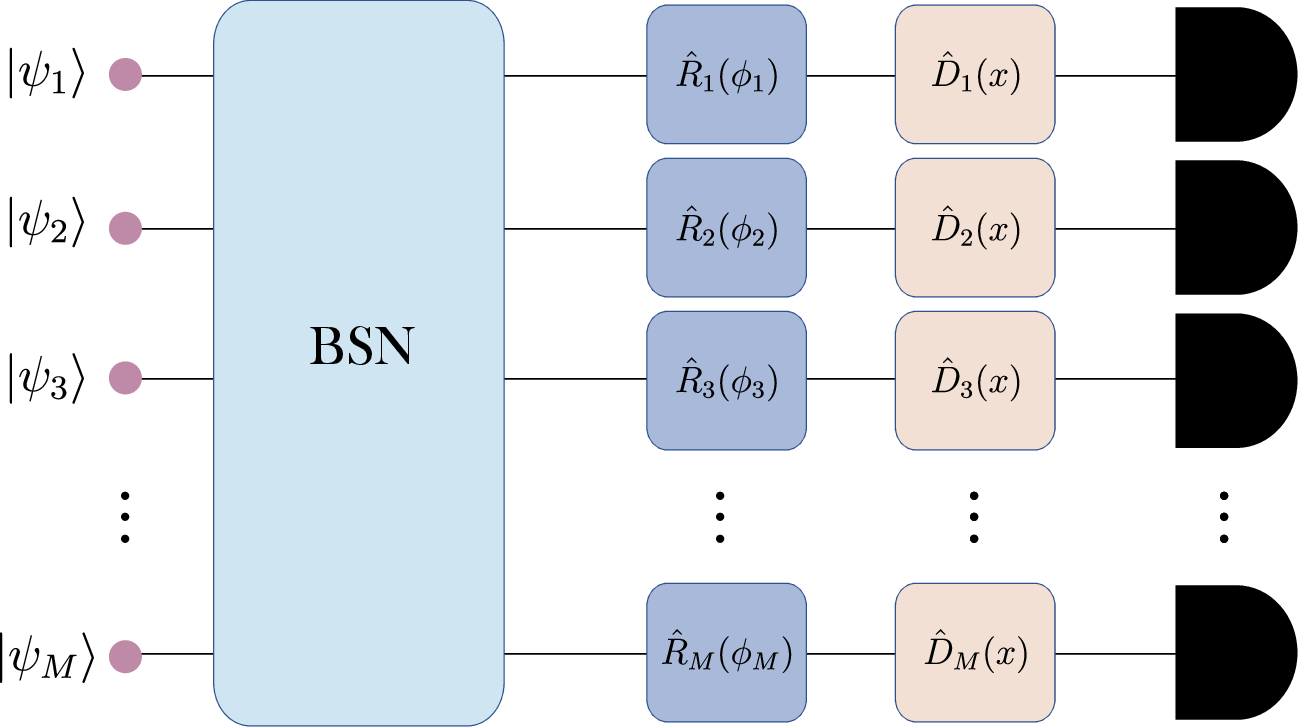}
    \caption{Schematic diagram of distributed quantum displacement sensing (see the main text).} 
    \label{fig:Circuit}
\end{figure}

\emph{Distributed quantum displacement sensing.---}
%We introduce a distributed quantum displacement sensing scheme using a CV quantum network on an $M$-mode CV system~\cite{zhuang2018distributed}.
%Our goal is to show that typical CV quantum networks allow a quantum enhanced estimation for such a displacement estimation task. 
For distributed displacement sensing (see Fig.~\ref{fig:Circuit}), we first prepare a product state with a total mean photon number $\bar{N}$.
The state is then injected into a BSN $\hat{U}$ to generate an entangled probe between $M$ modes.
Here, a BSN is described by an $M \times M$ unitary matrix $U$, which transforms input annihilation operators $\{\hat{a}_i\}_{i=1}^M$ as $\hat{a}_i \to \hat{U}^{\dagger} \hat{a}_i\hat{U}=\sum_{j=1}^{M} U_{ij}\hat{a}_j$.
After the BSN, we perform local phase shift operations, $\hat{R}(\bm{\phi})\equiv \otimes_{j=1}^{M} \hat{R}_j(\phi_j)$ with $\hat{R}_j(\phi_j)\equiv e^{i{\phi_j}\hat{a}_j^\dagger\hat{a}_j}$ being a phase-shift operator on $j$th mode for $\phi_j$.
Thus, for a given BSN, a local-phase optimization is implemented by manipulating $\phi_j$'s.
The entangled probe then encodes a displacement parameter $x$ of interest.
We assume that the same displacement occurs in all $M$ modes, the operator of which is written as $\otimes_{j=1}^{M}\hat{D}_j(x)$ with $\hat{D}_j(x)\equiv e^{-i \hat{p}_{j}x}$ being a displacement operator for $j$th mode along $x$-direction.
Here, quadrature operators of $j$th mode are defined as $\hat{x}_{j} \equiv (\hat{a}_{j}+\hat{a}^{\dagger} _{j})/\sqrt{2}$, $\hat{p}_{j} \equiv (\hat{a}_{j}-\hat{a}^{\dagger} _{j})/\sqrt{2}i$ for $x$ and $p$ directions in phase space, respectively.
Finally, we locally measure the output state on each site using homodyne detection and estimate the parameter $x$ using the measurement outcomes.
%Figure \ref{fig:Circuit} illustrates our distributed displacement sensing scheme.
%\delete{We emphasize that in our scheme, an extra quantum network other than a BSN to generate an entangled probe is not allowed for measurement.} 
We emphasize that our scheme has tensor product inputs and local measurements, while only the BSN can generate entanglement.
Note that the proposed scheme is similar to the one in Ref.~\cite{zhuang2018distributed} except that we employ an arbitrary BSN instead of a balanced one.
Also, such a distributed sensing scheme can offer advantages for many quantum metrological applications~\cite{zhuang2018distributed,grosshans2002continuous,pirandola2008continuous,pirandola2015high}.

Meanwhile, when we estimate a parameter $\theta$ of interest using a quantum state probe $\hat{\rho}$, the estimation error of any unbiased estimator, $\Delta^2 \theta$, is bounded by the quantum Cram\'er-Rao lower bound as $\Delta^2 \theta\geq 1/H$, 
where $H$ is the quantum Fisher information (QFI) for a given system and a probe state $\hat{\rho}$~\cite{braunstein1994statistical, paris2009quantum}. 
Therefore, QFI quantifies the ultimate achievable estimation error using a given quantum state.
Especially for a pure state probe $|\psi\rangle$ and a unitary dynamics with a Hamiltonian operator $\hat{h}$, the QFI can be simplified as $H=4(\Delta^2\hat{h})_\psi$. %$H=4(\Delta^2\hat{h})_\psi\equiv 4(\langle\hat{h}^2\rangle_\psi-\langle\hat{h}\rangle_\psi^2)$.

%For distributed displacement sensing, the attainable \cor{estimation error without entanglement between modes decreases at most as $1/M$, such as} using a product of identical states for $M$ modes such as squeezed states~\cite{zhuang2018distributed}.

For a distributed displacement sensing, the  attainable QFI without an entangled probe scales at most linear in $\bar{N}$ and $M$ (e.g.,~a product of identical states for $M$ modes such as squeezed states)~\cite{zhuang2018distributed, supple}.
Remarkably, if one employs the optimal entangled scheme (see Eq.~\eqref{QFIMAX}), the corresponding QFI scales as $\bar{N}M$~\cite{zhuang2018distributed, supple}. 
Therefore, an entanglement provides an advantage for distributed quantum displacement sensing if one prepares a suitable CV quantum network, and the advantage is apparent from the scaling of $\bar{N}M$.
For the purpose of the Letter that is to study the scaling of QFI in terms of the number of sensor nodes, we inspect the behavior of QFI as the number of modes $M$ grows with fixing the mean photon number per mode $\bar{n}\equiv\bar{N}/M$. It is worth emphasizing that since random quantum networks do not evenly allocate the input energy, the number of photons occupying a single mode fluctuates and can be much larger than $\bar{n}$.

\emph{Results.---}
We first derive the QFI for distributed displacement sensing for a given CV quantum network, characterized by an $M\times M$ unitary matrix $U$, with a squeezed state input.
After a BSN and phase shifters, the probe state can be written as $|\psi\rangle=\hat{R}(\bm{\phi})\hat{U}|\psi_\text{\text{in}}\rangle$, where $|\psi_\text{\text{in}}\rangle$ is a product state of a squeezed state in the first mode and $(M-1)$ vacua in other modes.
Since the Hamiltonian operator is $\hat{h}=\sum_{j=1}^M \hat{p}_j$, the QFI for distributed displacement estimation can be obtained as
\begin{align} 
    H_{LO}(U)&=\max_{\bm{\phi}} 4(\Delta^2\hat{h})_\psi=2M+ 4\left(\sum_{a=1}^{M}\abs{U_{a1}}\right)^{2}f_{+}(\bar{n}M), \label{QFI} 
\end{align}
where we have defined $f_{+}(\bar{n}M)\equiv \bar{n}M+\sqrt{\bar{n}^2M^2+\bar{n}M}$.
Here, the optimality condition of local phases for a given $U$ is written as $e^{-i\phi^{*}_a} = U_{a1} / \abs{U_{a1}}$, which depends only on the first column of $U$.
It is worth emphasizing that the optimality condition is immediately obtained from $U$.
The derivation of the QFI and the optimality condition is provided in Ref.~\cite{supple}.

Since the factor $f_{+}(\bar{n}M)$ in Eq.~\eqref{QFI} is order of $M$ for fixed $\bar{n}$, whether the Heisenberg scaling can be achieved, i.e., $H_{LO}(U)\propto M^2$, is determined by the property of BSN $U$.
Particularly, for a trivial BSN, $U=\mathbb{1}_M$, we do not attain any entanglement and the QFI is linear in $M$.
Thus, it fails to achieve the Heisenberg scaling without entanglement.
Meanwhile, one may easily show that the QFI is maximized by a balanced BSN, i.e., $|U_{a1}|=1/\sqrt{M}$ for all $a$'s, which leads to the QFI as
\begin{align}
    H_{\text{max}}\equiv \max_U H_{LO}(U)=2M + 4Mf_{+}(\bar{n}M). \label{QFIMAX}
\end{align}
It clearly shows the quantum enhancement from an optimal CV quantum network and the entanglement generated from it.
One can also prove that $H_\text{max}$ is the maximum QFI not only in our scheme but also over any quantum states~\cite{supple}.

Since our goal is to show a quantum metrological enhancement of typical CV quantum networks, we now compute the average QFI over random CV quantum networks using Eq.~\eqref{QFI}, i.e., random unitary matrices drawn from the Haar measure $\mu$ on $U(M)$ group, and prove the following lemma (See Ref.~\cite{supple} for a proof):
\begin{lemma} \label{Lemma1}
The average QFI over random $U$ for distributed quantum displacement sensing using a single-mode squeezed state is 
\begin{align}\label{EQFI}
      \E_{U \sim \mu}\left[H_{LO}(U)\right] = 2M + 4\left[\frac{\pi}{4}(M-1)+1\right]f_{+}(\bar{n}M).
\end{align}
\end{lemma}
First of all, Lemma~\ref{Lemma1} shows that the average QFI over random CV quantum networks follows the Heisenberg scaling.
Also, note that for large $M$, the ratio of the average QFI to the maximum QFI $H_\text{max}$ approaches to $\pi/4$.
Therefore, one may expect that typical CV quantum networks render a quantum metrological advantage.
We prove that indeed, most CV quantum networks offer a quantum enhancement for estimating displacement.
\begin{theorem} \label{Th1}
For an $M$-mode CV quantum network, characterized by an $M\times M$ unitary matrix drawn from the Haar measure $\mu$ on the $M \times M$ unitary matrix group, the Heisenberg scaling of QFI can be achieved with a fraction of BSNs such that
\begin{align}\label{Th1eq}
   \Prob_{U \sim \mu} \left[ H_{LO}(U) = \Theta(M^2) \right] \geq 1 - \exp\left[-\Theta(M)\right].
\end{align}
\end{theorem}
{\it Proof sketch.} (See Ref.~\cite{supple} for a formal proof.)
From the concentration of measure inequality~\cite{oszmaniec2016random, meckes2019random}, we have
\begin{align} \label{concen}
\Prob_{U\sim \mu} \left[\abs{f(U)- \E_{U\sim \mu}\left[f(U)\right]} \geq \epsilon\right] \leq 2\exp\left(-\frac{M{\epsilon}^2}{4 L^2}\right),
\end{align}
where $f: U \mapsto \mathbb{R}$ is a real function and $L$ is its Lipschitz constant. 
If we let $f(U)\equiv H_{LO}(U)$, the average $H_{LO}(U)$ is given by Lemma~\ref{Lemma1}.
We then show that $L$ is upper-bounded by $8Mf_+(\bar{n}M)$.
Finally, setting $\epsilon=\Theta(M^2)$ leads to Eq.~(\ref{Th1eq})~\cite{bigO}.

Since a product state renders QFI at most linear in $M$, Theorem $\ref{Th1}$ indicates that typical CV quantum networks with a squeezed vacuum state are beneficial for quantum metrology.
In other words, for a randomly chosen CV quantum network except for an exponentially small fraction, the proposed scheme achieves the Heisenberg scaling of QFI for the distributed displacement estimation.
It also implies that most CV quantum networks enable one to construct an entanglement using a single-mode squeezed vacuum state since the Heisenberg scaling can only be achieved using entanglement in our scheme.
Moreover, we prove that the QFIs can always be attained by performing homodyne detection along $x$-axis without an additional network~\cite{supple}.
Since the input state is product and additional operations, such as local optimization and measurement, are local, the entanglement is constituted only from CV quantum networks.

While our scheme with a squeezed vacuum state at a fixed mode is sufficient for our goal, the input state can be further optimized in principle.
For example, one may use an optimal input mode for a squeezed vacuum state for a given BSN or a product of squeezed vacuum states as an input.

Furthermore, since we can achieve the Heisenberg scaling using the optimal local phase shifts $\bm{\phi}^*$, Theorem \ref{Th1} can be interpreted from a different aspect. 
From the perspective of active transformation, the local phase shift for $i$th mode $\hat{R}_i(\phi_i^*)$ transforms the quadrature operator $\hat{p}_i$ into $\hat{R}_i^\dagger(\phi_i^*)\hat{p}_i\hat{R}_i(\phi_i^*)=\hat{x}_i \sin \phi_{i}^* + \hat{p}_i \cos \phi_{i}^*$.
Thus, if we absorb the local phase shifters into displacement operators by the above transformation, Theorem~\ref{Th1} implies that the QFI of the state right after a BSN mostly follows the Heisenberg scaling with respect to a parameter $x$ generated by operators $\sum_{i=1}^M(\hat{x}_i\sin\phi_i^*+\hat{p}_i\cos\phi_i^*)$.
Consequently, we obtain the following corollary: 
\begin{corollary}
When a single-mode squeezed vacuum state undergoes a random BSN, most of the output states are beneficial for distributed quantum displacement sensing with a specific direction of displacement.
\end{corollary}
Therefore, most CV quantum networks render an entanglement that enables one to attain the Heisenberg scaling for particular metrological tasks.
Nevertheless, if we fix the direction of displacement of interest, we find that local optimization is essential for our protocol.
In fact, without local operation, i.e., $\phi_a=0$ for all $a$'s, we cannot attain the Heisenberg scaling even if the input state is chosen to be the optimal state that maximizes QFI for a given $U$.
%Let us denote the QFI of the optimal state as $\mathcal{H}(U)$. 

\begin{theorem}\label{Th2}
Without local operation, the fraction of random BSNs for which QFI attains Heisenberg scaling is almost zero even if we choose the optimal input state for a given $U$,
\begin{align}
   \Prob_{U \sim \mu} \left[ \mathcal{H}(U) = \Theta(M^2) \right] \leq  \exp\left[-\Theta(M)\right].
\end{align}
where $\mathcal{H}(U)$ is the QFI of the optimal state.
\end{theorem}
{\it Proof sketch.}
First, we find an upper bound of the QFI of the optimal state for a given $U$ without local optimization.
We then show that the upper bound scales as $M$ except for an exponential small fraction of $U$'s in $M$, which implies that the QFI scales at most linearly in $M$ except for an exponentially small fraction of $U$'s.
The detailed proof is provided in Ref.~\cite{supple}.

    %Blue dots represent $H_{LO}$, the QFI when we fix the input mode and optimize the local phase shifts. 
    %The red dots represent the QFI when $\phi_j=0$ for all $j$'s and the input mode of the single-mode squeezed vacuum state is optimized 

\begin{figure}[t]
    \centering
    \includegraphics[width=0.95\linewidth]{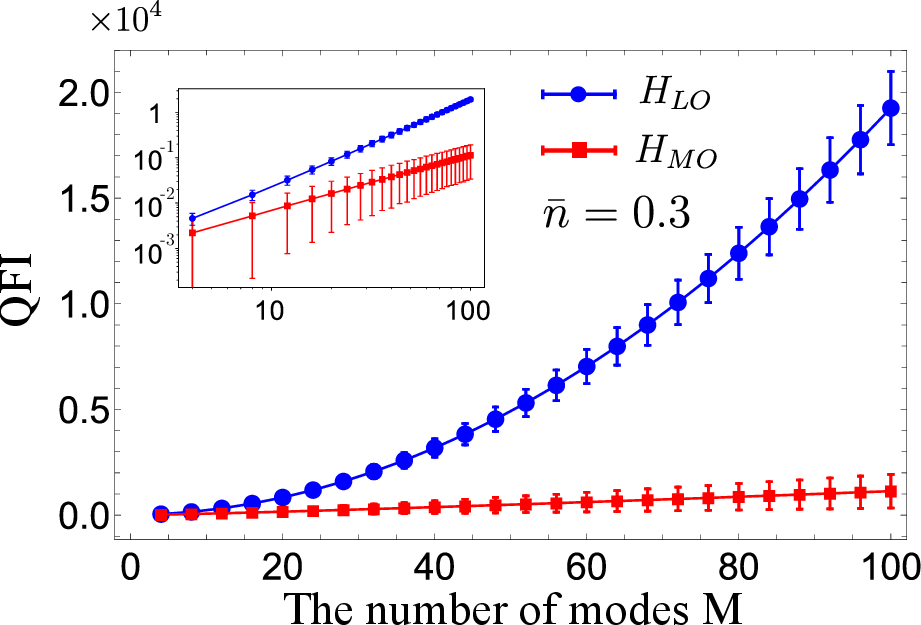}
    \caption{QFI averaged over 20000 different Haar-random BSNs with a squeezed state input
    (inset: log-log scale). 
    The error bars represent three times of the standard deviation of QFIs.
    }
    \label{fig:Haar}
\end{figure}

We now numerically demonstrate our results. We sample random unitary matrices by following the standard method that first generates Gaussian random matrix and orthogonalizes its column vectors~\cite{meckes2019random, supple}.
Figure \ref{fig:Haar} exhibits average QFIs over different Haar-random BSNs with a squeezed vacuum state input.
As implied by Theorems 1 and 2, it clearly shows that when we optimize the local phase shifts, we obtain QFIs following the Heisenberg scaling as the number of modes $M$ grows, while if we do not control the local phases, the Heisenberg scaling cannot be achieved~\cite{covariancedistance}.
Here, the QFI for a single-mode squeezed state input injected into an optimal input mode without local optimization is given by~\cite{supple}
\begin{align}
    H_{MO}\equiv \max_{1\leq b\leq M} \left[ 2M + 4\abs{\sum_{a=1}^{M}U_{ab}}^{2} f_{+}(\bar{n}M) \right]. \label{HMO}
\end{align}
Although we have used a single-mode squeezed state instead of the optimal input state, the overall scalings of $H_{MO}$ and $\mathcal{H}$ are equal when $M$ is large~\cite{supple}.
Furthermore, the standard deviation of QFIs are small for both cases, indicating that most BSNs with local-phase optimization allow the Heisenberg scaling using our scheme, while those without local-phase optimization does not.

\begin{figure}[t!]
    \centering
    \includegraphics[width=0.95\linewidth]{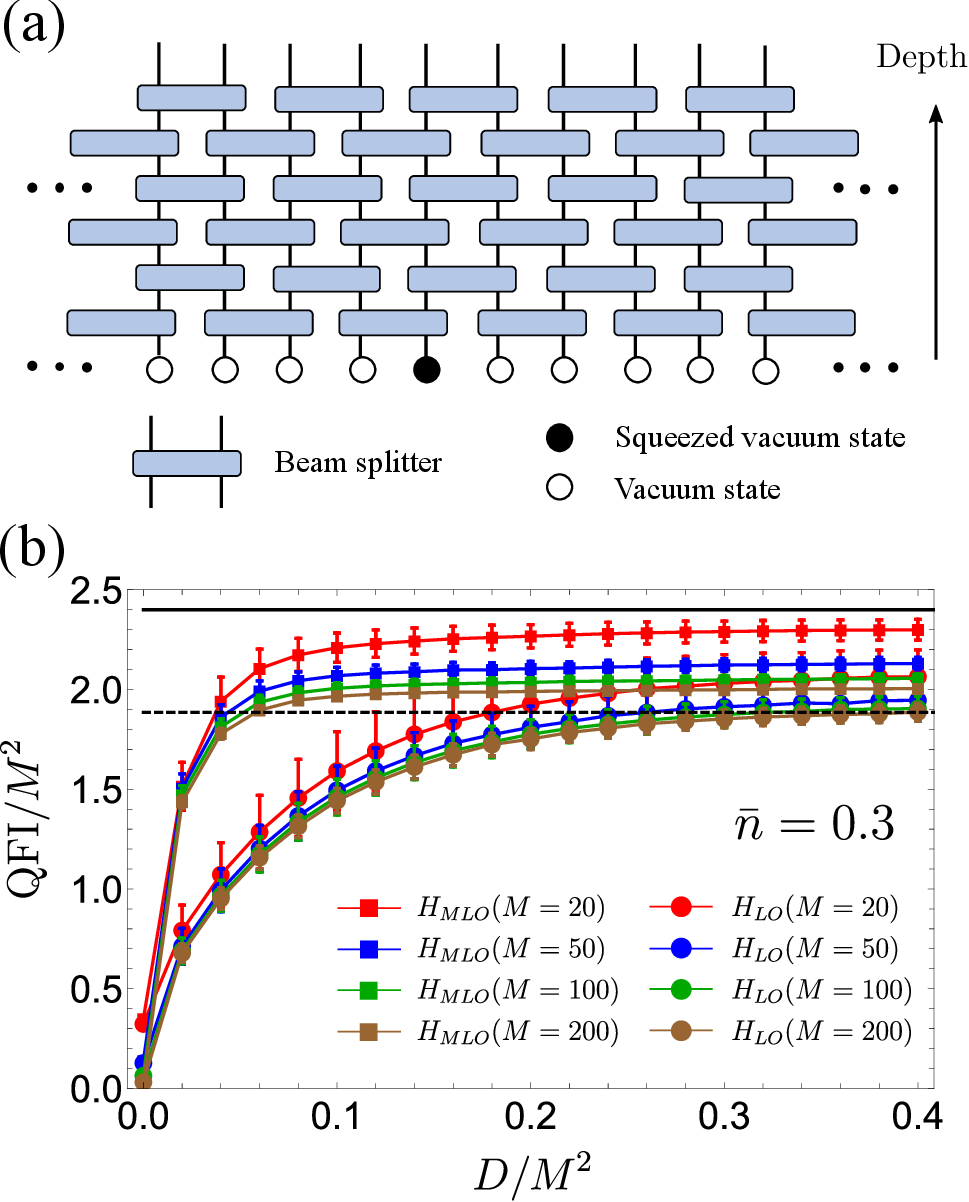}
    \caption{(a) CV quantum network composed of depth $D$ local beam splitters with a squeezed vacuum input. 
    %Here the depth $D$ is the number of layers of beam splitters. 
    %An input state is a squeezed vacuum state. 
    (b) Average QFIs over $1000$ different local Haar-random beam splitters with ($H_{MLO}$) and without ($H_{LO}$) optimizing the input mode. %different configurations, respectively. 
    %Average QFIs over different local Haar-random beam splitters for $M=20$ (red), $50$ (blue), $100$ (green), $200$ (brown) with ($H_{MLO}$) and without ($H_{LO}$) optimizing the input mode of the squeezed vacuum state. The average is taken over $10000, 1000, 1000, 500$ different configurations, respectively. 
    The error bars represent the standard deviation of QFIs over samples.
    Black dashed (solid) line represents the asymptotic average (maximum) QFI divided by $M^2$, obtained by a random (balanced) BSN, which is equal to $2\pi\bar{n}$ ($8\bar{n}$).}
    %Black dashed line represents the asymptotic value of the QFI divided by $M^2$ () obtained by random BSN and the single squeezed vacuum state on a fixed mode.} 
    \label{fig:local}
\end{figure}

\emph{Effect of loss.---}
We analyze the effect of photon loss on the Heisenberg scaling with typical BSNs and find a tolerable loss rate that maintains the Heisenberg scaling.
Photon loss can be modeled by a beam splitter with its transmittivity $\eta$, which transforms an annihilation operator as~$\hat{a}_j \rightarrow \sqrt{\eta} \hat{a}_j + \sqrt{1- \eta} \hat{e}_j$, where $\hat{e}_j$ is an annihilation operator for environment mode for all $j$'s~\cite{PhysRevA.48.3265}; thus, we assume that a photon-loss rate is constant over all modes. 
Since a photon-loss channel of the uniform loss rate commutes with beam splitters, our analysis includes photon loss occurring either before or after a BSN.
One can easily find that in the presence of photon loss, the corresponding QFI and its expectation value over random $U$ are degraded and their analytical expression can be written by merely replacing $f_{+}(\bar{n}M)$ in Eqs. \eqref{QFI} and \eqref{EQFI} by $\eta f_{+}(\bar{n}M)/\left[2(1-\eta)f_{+}(\bar{n}M)+1\right]$, which are shown in Ref.~\cite{supple}.
Using these results we can show that Theorem \ref{Th1} is still valid as long as a loss rate $1-\eta$ is smaller than a threshold $\beta=\Theta(1/\bar{n}M)$~\cite{supple}, i.e., as $M$ increases, a threshold of the loss rate has to decrease at least as $1/\bar{n}M$ to maintain the Heisenberg scaling. We note that CV error correction scheme~\cite{noh2020encoding, zhuang2020distributed} and quantum repeater~\cite{PhysRevA.99.012328} can be considered to alleviate the effect of loss.

%{Here we note that as $M$ increases, the tolerable threshold of the loss rate that enables to maintain the Heisenberg scaling decreases.}

\emph{Local beam splitter network.---}
While a global random BSN is suitable to model a sufficiently complex CV network, it is also crucial to investigate how complicated the network has to be to attain a metrological enhancement from a practical perspective.
To do that, we study a CV quantum network composed of local Haar-random beam splitters instead of a global random BSN (See Fig.~\ref{fig:local}(a))~\cite{zhuang2019scrambling,zhang2021entanglement,oh2021classical,oh2021classical2}.
We numerically show that the Heisenberg scaling can also be achieved by using CV quantum networks consisting of local beam splitters.
Figure \ref{fig:local}(b) shows the averaged local-phase-optimized QFIs with and without optimizing the input mode for a squeezed vacuum state.
The QFI of the latter is given by~\cite{supple}
\begin{align}
    H_{MLO}\equiv \max_{1\leq b\leq M} \left[ 2M + 4\left(\sum_{a=1}^{M}\abs{U_{ab}}\right)^{2} f_{+}(\bar{n}M) \right], \label{HMLO}
\end{align}
which is obviously equal or greater than $H_{LO}(U)$.
%Here, we have averaged the QFI over different quantum networks sampled from local Haar-random beam splitters.
Most importantly, the QFI divided by $M^2$ is almost constant for a given $D/M^2$ and different $M$'s.
It implies that the Heisenberg-scaling can be achieved on average with a depth proportional to $M^2$, independent of input-mode optimization, which is consistent with the result in Ref.~\cite{zhang2021entanglement}.
Nevertheless, by optimizing the input mode, the Heisenberg scaling is achieved much faster.
Moreover, the figure shows that the standard deviation of QFIs is very small, indicating that most local BSNs are beneficial for distributed displacement estimation, and that the standard deviation decreases as $M$ grows.
Since they achieve the Heisenberg scaling on average, the quantum networks of local beam splitters constitute sufficient entanglement on average as expected in Ref.~\cite{zhang2021entanglement}; namely, large entanglement can be obtained for a depth $D\propto M^2$.

\emph{Discussion.---}
%We have shown that typical CV quantum networks provide a quantum enhancement for distributed quantum displacement sensing.
From a theoretical perspective, our results imply that most CV quantum networks have the same scaling of estimation error for distributed displacement sensing as the optimal one, i.e., from a balanced BSN.
Thus, for quantum enhancement in practice, one may not necessarily implement a very special structure such as a balanced BSN because most CV networks provide the same quantum enhancement when it comes to scaling.
Such an experimental generalization would be particularly useful when one needs a large scale of networks.
For example, if we already have a CV quantum network for various purposes, which is not necessarily balanced but complex enough, we can immediately exploit the network for quantum-enhanced displacement sensing. 
Furthermore, although we have focused on distributed displacement sensing, future research could continue to investigate if similar results hold for different metrological tasks, such as multiparameter displacement estimation~\cite{zhuang2018distributed, xia2020demonstration} or phase estimation~\cite{ge2018distributed, guo2020distributed, oh2020optimal}.
It is also worth mentioning that since our scheme only employs a squeezed state, beam splitters, and homodyne detection, the current technology can already benefit from our results.

We finally emphasize the major differences of our study from Ref.~\cite{oszmaniec2016random}. 
While both consider random bosonic states from the quantum metrological perspective, the two schemes benefit from different kinds of entanglement. 
Ref.~\cite{oszmaniec2016random} studies phase sensing that exploits interparticle entanglement, while we study distributed displacement sensing which benefits from intermode entanglement. 
The difference is apparent from the following example.
The random state $\hat{R}(\bm{\phi}^*)\hat{U}\ket{N,0,\cdots,0}$ has mode entanglement and typically brings quantum enhancement for the distributed displacement sensing task whereas it has no particle entanglement regardless of BSN $U$, so it does not lead to an enhancement for phase estimation~\cite{supple}.
Besides, our study considers a task where a photon number fluctuates, which is unclear to interpret by particle formalism, typically assuming a definite photon number.
It would be an interesting future work to find a class of probes that is useful for both sensing schemes and to identify the relation between interparticle entanglement and intermode entanglement more rigorously.

%\cor{We finally emphasize the similarity and differences between our scheme and Ref.~\cite{oszmaniec2016random}'s. While both consider random bosonic states from the quantum metrological perspective, but deal withdifferent ensemble of the input state and different sensing schemes in terms of photon number preserving.  Specifically, Ref.~\cite{oszmaniec2016random} considers a quantum state which has a definite photon number and phase sensing scheme that preserves the photon number, in contrast, we allow an indefinite photon number state like a squeezed vacuum state and displacement sensing which does not preserve a photon number. Furthermore, the two different schemes benefit from different kinds of entanglement. Ref.~\cite{oszmaniec2016random} considers a phase sensing scheme that exploits particle entanglement, while we study a displacement sensing scheme benefits from mode entanglement (entanglement between the distant nodes). A quantum probe $\hat{R}(\bm{\phi}^*)\hat{U}\ket{N,0,\cdots,0}$~\cite{discussion} shows it explicitly. $\hat{R}(\bm{\phi}^*)\hat{U}\ket{N,0,\cdots,0}$ has a mode entanglement and brings quantum enhancement for distributed displacement sensing task but the probe does not have a particle entanglement so it cannot allow the enhancement for the phase estimation~\cite{supple}. It would be an interesting future work to find a class of probes that is useful for both sensing schemes.}

We acknowledge useful discussions with Quntao Zhuang.
H.K. and H.J. are supported by the National Research Foundation of Korea (NRF-2019M3E4A1080074, NRF-2020R1A2C1008609, NRF-2020K2A9A1A06102946) via the Institute of Applied Physics at Seoul National University and by the Ministry of Science and ICT, Korea, under the ITRC (Information Technology Research Center) support program (IITP-2021-2020-0-01606) supervised by the IITP (Institute of Information $\&$ Communications Technology Planning $\&$ Evaluation).
Y.L. acknowledges National Research Foundation of Korea a grant funded by the Ministry of Science and ICT (NRF-2020M3E4A1077861) and KIAS Individual Grant
(CG073301) at Korea Institute for Advanced Study.
L.J. and C.O. acknowledge support from the ARO (W911NF-18-1-0020, W911NF-18-1-0212), ARO MURI (W911NF-16-1-0349, W911NF-21-1-0325), AFOSR MURI (FA9550-19-1-0399), DoE Q-NEXT, NSF (EFMA-1640959, OMA-1936118, EEC-1941583, OMA-2137642), NTT Research, and the Packard Foundation (2013-39273).
We also acknowledge the University of Chicago’s Research Computing Center for their support of this work.

\bibliography{reference.bib}

\begin{thebibliography}{44}
\bibitem [{big()}]{bigO}%
  \BibitemOpen
  \href@noop {} {}\bibinfo {note} {We write $f(M)=\Theta(g(M))$ if $f(M)$ and
  $g(M)$ have the same behavior in the limit of large $M$.}\BibitemShut {Stop}%
\bibitem [{cov()}]{covariancedistance}%
  \BibitemOpen
  \bibinfo {note} {We also show in Ref. \cite{supple} that for large $M$, the
  average distance between the covariance matrices of random probes and the
  optimal probe converges to a nonzero constant. Interestingly, the average
  distance with local optimization is smaller than that without local
  optimization.}\BibitemShut {Stop}%
\end{thebibliography}
\end{document}

\definecolor{navy}{RGB}{46,72,102}
\definecolor{pink}{RGB}{255,249,249}
\definecolor{grey}{RGB}{184,184,184}
\definecolor{yellow}{RGB}{255,192,0}
\definecolor{grey1}{RGB}{217,217,217}
\definecolor{grey2}{RGB}{166,166,166}
\definecolor{grey3}{RGB}{89,89,89}
\definecolor{red}{RGB}{255,0,0}

\preprint{APS/123-QED}

\title{Supplemental Material: Quantum Metrological Power of Continuous-Variable Quantum Networks}
\author{Hyukgun Kwon}
\affiliation{Department of Physics and Astronomy, Seoul National University, Seoul 08826, Republic of Korea}
\author{Youngrong Lim}
\affiliation{School of Computational Sciences, Korea Institute for Advanced Study, Seoul 02455, Korea}
\author{Liang Jiang}
\affiliation{Pritzker School of Molecular Engineering, The University of Chicago, Chicago, Illinois 60637, USA}
\author{Hyunseok Jeong}
\email{h.jeong37@gmail.com}
\affiliation{Department of Physics and Astronomy, Seoul National University, Seoul 08826, Republic of Korea}
\author{Changhun Oh}
\email{changhun@uchicago.edu}
\affiliation{Pritzker School of Molecular Engineering, The University of Chicago, Chicago, Illinois 60637, USA}

\maketitle

\renewcommand{\theequation}{S\arabic{equation}}
\renewcommand{\thefigure}{S\arabic{figure}}
\renewcommand{\thesection}{S\arabic{section}}

\section{Optimal states for distributed displacement sensing} \label{Appen A}
In this section, our ultimate goal is to find the optimal quantum state that maximizes the QFI for distributed displacement sensing. 
In our scheme, an input state is an $M$-mode product state whose total mean photon number is $\bar{N}$, which we denote as $\ket{\psi_{\text{in}}} = \ket{\psi_1} \otimes \ket{\psi_2} \otimes \cdots \otimes \ket{\psi_M}$. After the state undergoes a BSN and phase shift operation, the state becomes
\begin{align}
    \ket{\psi_{\text{in}}} \to \ket{\psi}=\hat{R}(\phi) \hat{U} \ket{\psi_{\text{in}}}
\end{align}
where $\bm{\phi} = (\phi_1,\phi_2,\dots,\phi_M)$.
After these operations, displacement parameter $x$ is encoded on the state by the displacement operator $\hat{D}(x)   = \otimes_{j=1}^{M}e^{-i \hat{p}_{j}x} = e^{-i\hat{P}x}$ where $\hat{P}\equiv \hat{p}_1 + \hat{p}_2 + \cdots + \hat{p}_M$.  
Due to the facts that $\hat{R}(\phi) \hat{U} \ket{\psi_{\text{in}}}$ is a pure state and $\hat{D}(x)$ is an unitary operator, the QFI $H(U,\phi , \ket{\psi})$ is $4$ times of the variance of $\hat{P}$ \cite{liu2015quantum} :
\begin{align} \label{A2}
    H(U,\phi,\ket{\psi_{\text{\text{in}}}}) = 4 \left(\Delta^2 \hat{P} \right)_{\psi}=4\left(\langle\hat{P}^2\rangle_\psi-\langle\hat{P}\rangle_\psi^2\right).
\end{align}
To find the optimal states for distributed displacement sensing, first we focus on finding the single-mode state whose variance of $\hat{p}$ is the largest among all states having a mean photon number $\bar{n}$.
Using the Heisenberg uncertainty relation $\Delta^2 \hat{x}\Delta^2\hat{p} \geq 1/4$ and the mean photon number constraint $\bra{\Psi} \frac{1}{2}\left({\hat{x}}^{2}  +{\hat{p}}^{2} \right) \ket{\Psi} = \bar{n} + \frac{1}{2}$, we can derive the following inequality:
\begin{align} \label{A4}
    \bra{\Psi} {\hat{p}}^{2} \ket{\Psi} + \frac{1}{4\bra{\Psi} {\hat{p}}^{2} \ket{\Psi}} \leq 2 \bar{n} +1.
\end{align}
By simple calculation, we can get the maximum value of $\bra{\psi} {\hat{p}}^{2} \ket{\psi}$ which satisfies the equality in Eq. \eqref{A4}:
\begin{align} \label{A5}
    \max_{\ket{\Psi}}\bra{\Psi} {\hat{p}}^{2} \ket{\Psi} = \frac{2\bar{n} + 1 + 2\sqrt{{\bar{n}}^2 + \bar{n}}}{2}.
\end{align}

Meanwhile, one can easily check that the variance of a rotated quadrature operator $\hat{p}'=\hat{x}\sin \theta + \hat{p} \cos \theta$ of the $\ket{sqz(\theta,\bar{n})}$ of a single-mode squeezed vacuum state, defined as $\ket{sqz(\theta , \bar{n})}=\exp\left[\frac{1}{2}r\left( \hat{a}^{\dagger 2}e^{-2i\theta}- \hat{a}^{2}e^{2i\theta}\right) \right] \ket{0}$ whose mean photon number is $\bar{n}=\sinh^2 r$ \cite{agarwal2012quantum}.
\begin{align} \label{A3}
    \Delta^2 \hat{p}' = \bra{sqz(\theta,\bar{n})} {\hat{p}}^{\prime 2} \ket{sqz(\theta,\bar{n})} = \frac{e^{2r}}{2} = \frac{2\bar{n} + 1 + 2\sqrt{{\bar{n}}^2 + \bar{n}}}{2}.
\end{align}
By using Eq. \eqref{A3}, we find the optimal state. 
By comparing Eq. \eqref{A5} and \eqref{A3}, we can ensure that a squeezed vacuum state maximizes the variance $\Delta^2 \hat{p}$.

Using the above facts, we find the state that maximizes the QFI for estimating $x$, or equivalently the variance of $\hat{P}$. Before starting our main discussion, let us show how the quadrature operators $\left\{\hat{x}_i\right\}_{i=1}^{M}$ and $\left\{\hat{p}_i\right\}_{i=1} ^{M}$ transform via an $M$-mode BSN operator $\hat{U}$ and a local phase shift operator $\hat{R}(\bm{\phi})=\otimes_{j=1}^{M} \hat{R}_j(\phi_j)=\otimes_{j=1}^{M} \exp\left(i{\phi_j}\hat{a}_j^\dagger\hat{a}_j\right)$.

Second, let us show how the quadrature operators transform via a BSN and local phase shifts. An $M$-mode BSN operator $\hat{U}$ transforms the operators $\{\hat{a}_i\}_{i=1}^M$, where $\hat{a}_{i}$ represents annihilation operators for $i$th mode, as the following form:
\begin{align} \label{A6}
    \hat{a} _i  \to \hat{U}^{\dagger} \hat{a}_{i} {\hat{U}} =  \sum_{j=1}^{M} U_{ij}\hat{a}_j,
\end{align}
where $U$ is an $M \times M$ unitary matrix.
Using Eq. \eqref{A6}, we can find the transformations of the quadrature operators $\hat{x}_i= \left(\hat{a}_i + \hat{a}^{\dagger} _i \right)/ \sqrt{2}$ and $\hat{p}_i=\left(\hat{a}_i - \hat{a}^{\dagger}_i \right)/ i\sqrt{2}$ via $\hat{U}$:
\begin{align} 
    \hat{x}_i &\to  \hat{U}^{\dagger}\hat{x}_i {\hat{U}} = \sum_{j=1}^{M}\left[\left(\frac{U_{ij} + U_{ij}^{*}}{2}\right)\hat{x}_j - \left(\frac{U_{ij} - U_{ij}^{*}}{2i}\right)\hat{p}_j\right] = \sum_{j=1}^{M}\left(\hat{x}_j \Re U_{ij} -\hat{p}_j \Im U_{ij} \right)
    , \label{A7} \\
    {\hat{p}_i}&\to  \hat{U}^{\dagger} \hat{p}_i {\hat{U}} =\sum_{j=1}^{M}\left[\left(\frac{U_{ij} - U_{ij}^{*}}{2i}\right)\hat{x}_j + \left(\frac{U_{ij} + U_{ij}^{*}}{2}\right)\hat{p}_j \right] = \sum_{j=1}^{M}\left(\hat{x}_j \Im U_{ij} +\hat{p}_j \Re U_{ij} \right). \label{A8} 
\end{align} 
The transformations of quadrature operators via local phase shift operator are
\begin{align} 
    \hat{x}_i & \to  \hat{R}^{\dagger}(\bm{\phi}) \hat{x}_i \hat{R}(\bm{\phi}) = \hat{x}_i \cos \phi_{i} - \hat{p}_i \sin \phi_{i},  \label{A9} \\ 
    \hat{p}_i &\to  \hat{R}^{\dagger}(\bm{\phi}) \hat{p}_i \hat{R}(\bm{\phi}) = \hat{x}_i \sin \phi_{i} + \hat{p}_i \cos \phi_{i}. \label{A10}
\end{align}
Let us back to our main discussion. Noting that the mean photon number of the state is $\bar{N}$, we find an upper bound of the QFI as follows:
\begin{align}
      H(U,\phi,\ket{\psi_{\text{in}}})&=4\left(\Delta \hat{P}\right)^2\leq 4\bra{\psi_{\text{in}}} {\hat{U}}^{\dagger} \hat{R}^{\dagger}(\bm{\phi}){\hat{P}}^{2} \hat{R}(\bm{\phi}) \hat{U} \ket{\psi_{\text{in}}}=4\bra{\psi_{\text{in}}} \left[\sum_{a=1}^{M}\sum_{b=1}^{M} \hat{x}_b \left(\Im e^{i\phi_a}U_{ab} \right)  + \hat{p}_b \left(\Re e^{i\phi_a} U_{ab}\right)   \right]^{2} \ket{\psi_{\text{in}}} \label{A11}\\ 
     &= 4\sum_{b=1}^{M}\bra{\psi_b} \left[{\hat{x}_b}^{2} \left(\sum_{a=1}^{M} \Im e^{i\phi_a} U_{ab}\right)^2 + {\hat{p}_b}^{2}\left(\sum_{a=1}^{M} \Re e^{i\phi_a}  U_{ab}\right)^{2}\right] \ket{\psi_b} \label{A12}\\ 
     &=  4\sum_{b=1}^{M} R_{b} \bra{\psi_b} {\hat{x}_b}^{2} \sin^2 \theta_b  + {\hat{p}_b}^{2} \cos^2 \theta_b  \ket{\psi_b}
     =  4\sum_{b=1}^{M} R_{b} \bra{\psi_b} \left({\hat{x}_b} \sin \theta_b  +   {\hat{p}_b} \cos \theta_b\right)^{2} \ket{\psi_b} \\ 
     &=  4\sum_{b=1}^{M} R_{b} \bra{\psi_b} {\hat{p}^{\prime 2} _b} \ket{\psi_b} \leq 2\sum_{b=1}^{M} R_{b} \left(2\bar{n}_b + 1 + 2\sqrt{{\bar{n}_b}^2 + \bar{n}_b}\right) = 2\sum_{b=1}^{M} \abs{\sum_{a=1}^{M} e^{i\phi_a} U_{ab}}^{2} \left(2\bar{n}_b + 1 + 2\sqrt{{\bar{n}_b}^2 + \bar{n}_b}\right), \label{A14} 
\end{align}

where $\bar{n}_b$ denotes the mean photon number of mode $b$, $R_{j}(\bm{\phi})\equiv\left(\sum_{i=1}^{M} \Im e^{i\phi_i} U_{ij}\right)^{2} + \left(\sum_{i=1}^{M} \Re e^{i \phi_i} U_{ij}\right)^{2}$, $\left(\sum_{i=1}^{M} \Re e^{i\phi_i} U_{ij}\right)^{2}  R_{j}(\bm{\phi}) \equiv \cos ^2 \theta_j$ and $\left(\sum_{i=1}^{M} \Im e^{i\phi_i} U_{ij}\right)^{2}  R_{j}(\bm{\phi}) \equiv \sin ^2 \theta_j$. 
To get the equalities in Eqs. \eqref{A11}-\eqref{A14}, one can refer the Eqs. \eqref{A5}-\eqref{A3}. 
If the input state $\ket{\psi_{\text{in}}} = \ket{\psi_1} \otimes \ket{\psi_2} \otimes \cdots \otimes \ket{\psi_M} $ is $\ket{sqz(\theta_1,\bar{n}_{1})}\otimes\ket{sqz(\theta_2,\bar{n}_{2})} \otimes \cdots \otimes \ket{sqz(\theta_M,\bar{n}_{M})}$, the inequalities are saturated.
From now on, we call the state that saturates the inequalities, as a proper squeezed state.
Moreover, by using Eq. \eqref{A14}, we can find an upper bound of the $H$ for a given $U$. 
Particularly, once we choose an input state as a proper squeezed state, further optimization can be performed to maximize the $H$ by adjusting local phase shifts $\phi_a$s and photon number allocation $\bar{n}_b$s. 
Therefore, we can rewrite $H$ as a function of $U$, $\bm{\phi}$ and $\bar{\bm{n}}=(\bar{n}_1,\dots,\bar{n}_M)$:
\begin{align} \label{A15}
    H(U,\bm{\phi},\bar{\bm{n}})= 2M + 4\sum_{b=1}^{M} \abs{\sum_{a=1}^{M} e^{i\phi_a} U_{ab}}^{2} \left(\bar{n}_b + \sqrt{{\bar{n}_b}^2 + \bar{n}_b}\right) = 2M + 4M\sum_{b=1}^{M} p_b(\phi) f_{+}(\bar{n}_b),
\end{align}
where $p_b (\phi) \equiv \abs{\sum_{a=1}^{M} e^{i\phi_a} U_{ab}}^{2}/M$ and $f_{+}(x)\equiv x+\sqrt{x^2 +x}$. Note that $\sum_{b=1}^{M}\abs{\sum_{a=1}^{M} e^{i\phi_a} U_{ab}}^{2}=M$. 
To find an upper bound of $H$, we find an upper bound of $\sum_{b=1}^{M} p_b(\phi) f_{+}(n_b )$. 
Once we consider $\{p_b (\phi)\}_{b=1} ^{M}$ as a probability distribution, (note that $0 \leq p_b \leq 1$ for all $b$) we can derive following inequalities :
\begin{align}
    \sum_{b=1}^{M} p_b (\phi) f_{+}(\bar{n}_b ) \leq f_{+}\left(\sum_{b=1}^{M} p_b (\phi) \bar{n}_b \right) \leq f_{+}\left(\max\sum_{b=1}^{M} p_b (\phi) \bar{n}_b \right).
\end{align}
By using the fact that $f_{+}(x)$ is a concave and increasing function, the first and second inequality are established. 
The last inequality is saturated by the following $\bar{\bm{n}}$. 
Using the photon-number constraint $\sum_{b=1}^{M} \bar{n}_b =\bar{N}$, 
\begin{align}
    \sum_{b=1}^{M} p_b (\phi) \bar{n}_b
    =p_{\nu} (\phi)\left(\bar{N}-\sum_{b\neq \nu}^M\bar{n}_b \right) + \sum_{b\neq {\nu}}^{M} p_b (\phi) \bar{n}_b
    =p_{\nu}(\phi)\bar{N}+\sum_{b\neq {\nu}}^{M} p_b (\phi) (\bar{n}_b-\bar{n}_\nu)
    \leq p_\nu(\phi) \bar{N},
\end{align}
where we have chosen $\nu$ to be the index that corresponds to the maximum of $\{p_b (\phi)\}_{b=1}^M$. 
Consequently, the ultimate upper bound of $H$ is
\begin{align} \label{A18}
    H(U,\bm{\phi},\bar{\bm{n}})= 2M + 4\sum_{b=1}^{M} \abs{\sum_{a=1}^{M} e^{i\phi_a} U_{ab}}^{2} f_{+}(\bar{n}_b) \leq 2M + 4Mf_{+}(p_{\nu}(\bm{\phi})\bar{N})  \leq 2M + 4Mf_{+}(\bar{N}) \equiv H_{max}. 
\end{align}
Notice that all the inequalities in Eq. \eqref{A18} are saturated only when $p_{\nu}(\bm{\phi})$ is $1$ which is the case when the BSN is balanced. 
A balanced BSN satisfies $U_{a\nu}=1/\sqrt{M}$ for all $a$'s for some $\nu$. Note that Ref.~\cite{zhuang2018distributed} has shown that when a balanced BSN is used, a single-mode squeezed input state gives optimal estimation error for a particular measurement and estimator.
In contrast, we have not assumed any type of measurement and estimator by considering QFI and obtained the QFI in Eq.~\eqref{A18} which gives the same estimation error as in Ref.~\cite{zhuang2018distributed}.
It suggests that the proposed scheme in Ref.~\cite{zhuang2018distributed} is indeed optimal for distributed displacement sensing.

Using \eqref{A18}, we can also find the optimal product state without an entanglement. Let us consider the trivial BSN which satisfies $U_{ab}=\delta_{ab}$. In this case, the QFI satisfies following inequality:
\begin{align} \label{A19}
    H(U=\mathbb{1}_{M},\bm{\phi},\bar{\bm{n}})= 2M + 4\sum_{b=1}^{M}f_{+}(\bar{n}_b) \leq 2M + 4 \sqrt{\sum_{b=1}^{M}1}\sqrt{\sum_{b=1}^{M}f^{2}_{+}(\bar{n}_b)}.
\end{align}
The inequality in \eqref{A19} comes from the Cauchy-Schwarz inequality and saturated only when $\bar{n}_b = \bar{n}$ for all $b$. This implies that the optimal product state without an entanglement for distributed displacement sensing is a tensor product of identical squeezed states and the corresponding QFI is $2M+4\sum_{b=1}^{M}f_{+}(\bar{n})$.

\section{Local optimized QFI}\label{Appen B}
The first step to prove Lemma 1 and Theorem 1 is to find the local-phase-optimized QFI $H_{LO}(U)$ in Eq. (2).
In our scheme, we consider a single-mode squeezed vacuum state as an input state, squeezed along the $x$ axis with mean photon number $\bar{N}$. 
More specifically, the squeezed state is injected into the first mode and other modes are in the vacuum regardless of $U$. 
(This state might not be a proper squeezed state.) 
The state undergoes a BSN, local phase shift operations, and displacement encoding. 
The corresponding QFI can be found via Eq. \eqref{A2} : 
\begin{align} 
    H(U, \bm{\phi}) &=2M+ 4\left[\left(\Re\sum_{a=1}^{M}e^{i\phi_a}U_{a1}\right)^{2} f_{+}(\bar{N})  + \left(\Im\sum_{a=1}^{M}e^{i\phi_a}U_{a1}\right)^{2} f_{-}(\bar{N})\right] \label{B1} 
\end{align}
where $f_{-}(x)=x-\sqrt{x^2+x}$. We emphasize again that $H_{LO}(U)$ is the optimized QFI via local phase shift operations, i.e., $H_{LO}(U) \equiv \max_{\bm{\phi}}H(U, \bm{\phi})$. We can find an upper bound of $H(U, \bm{\phi})$ by using following inequalities :
\begin{align} 
    H(U, \bm{\phi}) &=2M+ 4\left[\left(\Re\sum_{a=1}^{M}e^{i\phi_a}U_{a1}\right)^{2} f_{+}(\bar{N})  + \left(\Im\sum_{a=1}^{M}e^{i\phi_a}U_{a1}\right)^{2} f_{-}(\bar{N})\right] \leq 2M + 4\left(\Re\sum_{a=1}^{M}e^{i\phi_a}U_{a1}\right)^{2} f_{+}(\bar{N}) \nonumber \\
    &\leq 2M + 4\abs{\sum_{a=1}^{M}e^{i\phi_a}U_{a1}}^{2} f_{+}(\bar{N}) \leq 2M + 4\left(\sum_{a=1}^{M}\abs{U_{a1}}\right)^{2} f_{+}(\bar{N}).
\end{align}
The first inequality holds because $f_{-}(N) \leq 0$ and others are straightforward to derive. Here, all of the above inequalities are saturated when all the $\phi_a$'s satisfy the condition $e^{-i\phi^{*}_a}=U_{a1}/{\abs{U_{a1}}}$. Finally, we get the local optimized QFI:
\begin{align} 
  H_{LO}(U)\equiv \max_{\bm{\phi}}H(U,\bm{\phi})=2M + 4\left(\sum_{a=1}^{M}\abs{U_{a1}}\right)^{2} f_{+}(\bar{N}). \label{B4}
\end{align}
Note that $H_{LO}$ is not fully optimized QFI over input states. 
Even if we only consider the input state as a single mode squeezed vacuum state, the QFI can be further optimized by choosing the optimal input mode depending on a given BSN instead of injecting the state into the first mode. 
When a squeezed vacuum state is injected into the $b$th mode, the corresponding QFI is given by $2M + 4\left(\sum_{a=1}^{M}\abs{U_{ab}}\right)^{2} f_{+}(\bar{N})$. Therefore, more optimized QFI over both input mode and local phase is written as
\begin{align}
H_{MLO}(U) \equiv \max_{1\leq b\leq M} \left[2M + 4\left(\sum_{a=1}^{M}\abs{U_{ab}}\right)^{2} f_{+}(\bar{n}M)\right],
\end{align}
which is always equal or greater than $H_{LO}(U)$.

\section{Proof of Lemma 1 and Theorem 1} \label{Appen C}
\subsection{Proof of Lemma 1}
Before presenting the proof of Lemma 1, we briefly introduce one of the methods to generate Haar-random unitary matrix on $U(M)$ group using \textit{Ginibre} ensemble and \textit{Gram-Schmidt} orthogonalization. Suppose there is a random $M \times M$ matrix $Z$ whose components $z_{ij}$s are mutually independent and each of the entry follows the standard complex normal distribution $P(z_{ij})d^{2}z_{ij}=\frac{1}{\pi}e^{-{\abs{z_{ij}}}^2}d^{2}z_{ij} = \frac{1}{\pi}e^{-{R_{ij}}^2}R_{ij}(dR_{ij})(d\theta_{ij})$ where $R=\abs{z_{ij}}$ and $\theta_{ij}$ is argument of $z_{ij}$. (This set of random matrices are called \textit{Ginibre} ensemble, or Gaussian random matrices.)
By performing the \textit{Gram-Schmidt} orthogonalization procedure as follows, one can generate $M \times M$ Haar-random unitary matrix \cite{meckes2019random}.
\begin{align} \label{C1}
  {U}_1 =\frac{{Z}_1}{\| {Z}_1\|}, \quad  {U}_2 = \frac{{Z}_2 - \left<{Z}_2 , {U}_1\right>{U}_1}{\| {Z}_2 - \left<{Z}_2 , {U}_1\right>{U}_1\|} , \quad {U}_3 = \frac{{Z}_3 - \left<{Z}_3 , {U}_1\right>{U}_1 - \left<{Z}_3 , {U}_2\right>{U}_2}{\|{Z}_3 - \left<{Z}_3 , {U}_1\right>{U}_1 - \left<{Z}_3 , {U}_2\right>{U}_2\|}, \quad \dots,
\end{align}
where $U_{i}$ and $Z_{i}$ are $i$th column vector of matrix $U$ and $Z$ each. Using Eq. \eqref{C1}, we can express $U_{a1}$ as $z_{a1}/\sqrt{\sum_{a=1}^{M}z_{a1}z^{*}_{a1}}$ or $R_{a1}e^{i\theta_{a1}}/\sqrt{\sum_{a=1}^{M}R^{2}_{a1}}$.

{\it Proof of Lemma 1.}
Our goal is to find the expectation value of local optimized QFI, which is simplified as
\begin{align} \label{C2}
    \E_{U\sim\mu}\left[H_{LO}(U)\right] = \E_{U\sim\mu}\left[2M + 4\left(\sum_{a=1}^{M}\abs{U_{a1}}\right)^{2} f_{+}(\bar{N})  \right] = 2M + 4f_{+}(\bar{N})\sum_{a,b=1}^{M}\E_{U\sim\mu}\left[\abs{U_{a1}}\abs{U_{b1}}\right].
\end{align}
Note that the expectation value over Haar-random only cares the term $\sum_{a,b=1}^{M}\abs{U_{a1}}\abs{U_{b1}}$.
Due to the procedure introduced above, we can simplify the term as
\begin{align} 
  \sum_{a,b=1}^{M}\E_{U\sim\mu}\left[\abs{U_{a1}}\abs{U_{b1}}\right]
  &= \left(\prod_{i=1}^{M} \int_{0}^{\infty} 2R_i dR_i e^{-{R^2 _i}}\int_{0}^{2\pi} \frac{d\theta_i}{2\pi}\right)\left(\sum_{a,b=1}^{M}\frac{R_a R_b }{\sum_{k=1}^{M}{R_k}^2}\right) \\
   &=\left(\prod_{i=1}^{M} \int_{0}^{\infty} 2R_i dR_i e^{-{R^2 _i}}\right)\left[\sum_{a=1}^{M}\left(\frac{R^{2} _a }{\sum_{k=1}^{M}{R_k}^2}\right) + \sum_{a\neq b}^{M}\left(\frac{R_a R_b }{\sum_{k=1}^{M}{R_k}^2}\right)\right] \label{C4} \\
   &= 1+ M(M-1)\left(\prod_{i=1}^{M} \int_{0}^{\infty} 2R_i dR_i e^{-{R^2 _i}}\right)\left(\frac{R_1 R_2 }{\sum_{k=1}^{M}{R_k}^2}\right) \label{C5} \\ &=1+ M(M-1)\left(\prod_{i=1}^{M-1}  \int_{0}^{\frac{\pi}{2}} d\phi_i \right)\left(\cos\phi_1 \sin\phi_1 \cos\phi_2\right)\left[\prod_{k=1}^{M-1} \left(\sin\phi_k\right)^{2M-2k-1} \cos\phi_k \right] \nonumber \\
  & ~~~~~~~~~~~\times\int_{0}^{\infty} dR 2^{M} (R)^{2M-1}e^{-R^2} \label{C6} \\
  &= 1 + \frac{\pi}{4}\left(M-1\right). \label{C7}
\end{align}
The equality between Eq. \eqref{C4} and \eqref{C5} holds because of the symmetry of $a,b$. In Eq. \eqref{C6}, we adopt  $M$-dimensional spherical coordinate \cite{blumenson1960derivation}. We express the integral variables $\{R_{a}\}_{a=1}^{M}$ as $R_1 = \left(R \cos \phi_1\right)$, $R_2 = \left(R \sin\phi_1 \cos \phi_2\right)$, $\dots$, $R_{M-1}=\left(R \sin\phi_1 \sin\phi_2 \cdots \sin\phi_{M-2} \cos \phi_{M-1}\right)$, $R_{M}=\left(R \sin\phi_1 \sin\phi_2 \cdots \sin\phi_{M-2} \sin \phi_{M-1}\right)$
and corresponding Jacobian determinant is $\prod_{i=1}^{M} dR_i = R^{M-1} \prod_{k=1}^{M-2} (\sin\phi_k)^{^{M-k-1} }$. 
Finally, the value in Eq. \eqref{C7} is deduced by the following integral table:
\begin{align} 
    &\int_{0}^{\frac{\pi}{2}}(\sin\phi)^{2M-2k-1} (\cos\phi) d\phi = \frac{1}{2} \left(\frac{1}{M-k}\right)~~~ \text{if}~~ M>k, \label{e31}\\ 
   &\int_{0}^{\frac{\pi}{2}}(\sin\phi)^{2M-2} (\cos\phi)^2 d\phi = \frac{\sqrt{\pi}}{4} \frac{\Gamma\left(\frac{2M-1}{2}\right)}{\Gamma\left(M+1\right)}~~~ \text{if}~~ M>\frac{1}{2}, \label{e32} \\
   &\int_{0}^{\frac{\pi}{2}}(\sin\phi)^{2M-5} (\cos\phi)^2 d\phi =  \frac{\sqrt{\pi}}{4} \frac{\Gamma\left(M-2\right)}{\Gamma\left(\frac{2M-1}{2}\right)}~~~ \text{if}~~ M>2,\label{e33} \\
   &\int_{0}^{\infty} R^{2M-1} e^{-R^2} dR = \frac{\Gamma(M)}{2}~~~ \text{if}~~ M>0, \label{e34}
\end{align}
where $\Gamma(\cdot)$ is the gamma function.

Hence, we complete the proof of Lemma 1 :
\begin{align}
    \E_{U\sim\mu}\left[H_{LO}(U)\right]=2M + 4\left[\frac{\pi}{4}(M-1)+1\right]f_{+}(\bar{N})=2M + 4\left[\frac{\pi}{4}(M-1)+1\right]\left(\bar{n}M + \sqrt{{\bar{n}}^{2}M^{2}+\bar{n}M}\right).
\end{align}
\subsection{Upper bound of Lipschitz Constant $L_O$ of $H_{LO}(U)$}
In this subsection we derive the upper bound of Lipschitz constant $L_{O}$ of $H_{LO}(U)$ with the aid of a function $H_{1}(U)$, which we introduce below. 
The distance between two unitary matrices $U$ and $U'=\exp\left[-iX\phi\right]U$ can be written as $\phi\|X\|_{HS}$ \cite{meckes2019random, oszmaniec2016random}. Here $X$ is an $M \times M$ Hermitian matrix and $\|X\|_{HS}\equiv\sqrt{\Tr\left[X^{\dagger}X\right]}$. Therefore the Lipschitz constant $L_f$ of a smooth function $f:\mathbb{U}(M) \mapsto \mathbb{R}$ is the smallest positive constant satisfying the following inequality \cite{meckes2019random, oszmaniec2016random}: 
\begin{align}
   \abs{\frac{d}{d\phi}f(e^{-iX\phi}U)}_{\phi=0} \leq L_{f}  \|X\|_{HS}. \label{C13}
\end{align}

Let us introduce a function $ H_{1}(U) \equiv 2M + 4 \abs{\sum_{a=1}^{M} U_{a1}}^{2} f_{+}(\bar{N}) $ and corresponding Lipschitz constant $L_{1}$. Note that $H_{1}(U)$ is the QFI when the input state is same with $H_{LO}(U)$'s but without local operation. Exploiting Eq. \eqref{C13}, we find the upper bound of $L_{1}$:
\begin{align}
  \abs{\frac{d}{d\phi}H_{1}(e^{-iX\phi}U)}_{\phi=0} &=  4f_{+}(\bar{N})\left[\left(\sum_{a=1}^{M}\sum_{d=1}^{M}-iX_{ad} U_{d1} \right)\left(\sum_{a^{^{\prime}}=1}^{M} {U^{*}_{a^{^{\prime}}1}} \right) +  \left(\sum_{a=1}^{M}\sum_{d=1}^{M}i{X^{*}_{ad}} {U^{*}_{d1}} \right)\left(\sum_{a^{^{\prime}}=1}^{M} U_{a^{^{\prime}}1} \right) \right]\\
  &= 8f_{+}(\bar{N}) \left[\sum_{d=1}^{M} \sum_{a=1}^{M} \Im\left(Y_d U_{d1} {U^{*}_{a1}} \right)  \right] 
  \leq 8f_{+}(\bar{N}) \left(\sum_{d=1}^{M}  \abs{Y_d} \abs{U_{d1}}\right) \left(\sum_{a=1}^{M} \abs{{U_{a1}}}  \right) \label{C15} \\
  &\leq 8f_{+}(\bar{N}) \sqrt{\sum_{d=1}^{M}  {\abs{Y_d}}^{2} \sum_{d=1}^{M} {\abs{U_{d1}}}^{2} } \sqrt{\sum_{a=1}^{M} {\abs{{U_{a1}}}}^{2} \sum_{a=1}^{M} 1 } = 8f_{+}(\bar{N}) \sqrt{\sum_{d=1}^{M}  {\abs{Y_d}}^{2}} \sqrt{M} \leq 8f_{+}(\bar{N})M \|X\|_{HS}  \label{C16}
\end{align}
where $Y_d \equiv \sum_{a=1}^{M} X_{ad}$. The first inequality in Eq. \eqref{C16} holds from the Cauchy-Schwarz inequality. The last inequality in Eq. \eqref{C16} holds by the following fact: 
\begin{align*}
   \sqrt{\sum_{d=1}^{M}  {\abs{Y_d}}^{2}} &
   = \sqrt{\sum_{d=1}^{M} \left(X_{1d} + X_{2d} + \dots + X_{Md}\right)\left(X_{1d}^{*} + X_{2d}^{*} + \dots + X_{Md}^{*}\right)}\\
   &=\sqrt{M\left(\sqrt{\frac{1}{M}},\sqrt{\frac{1}{M}},\sqrt{\frac{1}{M}},\dots,\sqrt{\frac{1}{M}}\right)X X^{\dagger}\left(\sqrt{\frac{1}{M}},\sqrt{\frac{1}{M}},\sqrt{\frac{1}{M}},\dots,\sqrt{\frac{1}{M}}\right)^{\intercal}} \leq \sqrt{M} \|X\| \leq  \sqrt{M} \|X\|_{HS},
\end{align*}
where $\|X\| \equiv \sup_{\ket{\psi}} \frac{\|X \ket{\psi}\|}{\| \ket{\psi}\|}$. We use the fact  $\|X\| \leq \|X\|_{HS}$ \cite{oszmaniec2016random}. Finally, we can conclude that the upper bound of $L_{1}$ is $8Mf_{+}(\bar{N})$. 

Using the upper bound of $L_{1}$ and the relation between $H_{LO}(U)$ and $H_{1}(U)$, we can find the upper bound of $L_{O}$. $H_{LO}(U)$ can be expressed by $H_{1}$ as  $H_{LO}(U)=H_{1} \left(VU\right)$ where $V= \text{diag}\left(U^{*}_{11}/\abs{U_{11}}, U^{*}_{21}/\abs{U_{21}}\cdots  U^{*}_{M1}/\abs{U_{M1}}\right)$ is the unitary matrix corresponding to the optimal phase shifter. 
Note that $H_{1}(U) \leq H_{LO}(U)$ for all $U$. Without loss of generality, assume that $H_{LO}(U') \geq H_{LO}(U)$. We then derive the following equations:
\begin{align}
    \abs{H_{LO}(U') - H_{LO}(U)} = H_{LO}(U') - H_{LO}(U)=H_{1}(V'U') - H_{1}(VU)  \leq H_{1}(V'U') - H_{1}(V'U) .
\end{align}
The last inequality comes from the fact that $H_{1}(VU)\geq H_{1}(V'U)$. 
The above inequalities imply that the upper bound of $L_{O}$ is $L_{1}$ because the distance between $V'U'$ and $V'U$ is the same as the one between $U'$ and $U$. 
Therefore we can find that $L_{O}$ is bounded as $L_{O} \leq 8Mf_{+}(\bar{N})$.

\subsection{Proof of Theorem 1}
Before proceeding the proof of Theorem 1, let us introduce concentration of measure inequality \cite{oszmaniec2016random,anderson2010introduction}. For a smooth function $f: U \mapsto \mathbb{R}$ where $U$ is drawn from Haar measure on $M \times M$ unitary matrix group $\mu$, the following inequality holds \cite{anderson2010introduction}:
\begin{align}
      \Prob_{U\sim\mu}\left(\abs{ f(U) - \E_{U \sim \mu}[f(U)]} \geq \epsilon\right)  
      \leq 2\exp\left(-\frac{M {\epsilon}^2}{4L^{2}_{f}}\right), \label{C18}
\end{align}
where $L_{f}$ is Lipschitz constant of $f(U)$. 

In Theorem 1, we claim that most of the local-phase-optimized QFI attains Heisenberg scaling. We process Eq. \eqref{C18} to prove our claim: 
\begin{align}
    \Prob_{U\sim\mu}\left(H_{LO}(U) \leq (2\pi\bar{n}-k)M^{2}\right) &\leq
    \Prob_{U\sim\mu}\left(H_{LO}(U) \leq \E_{U \sim \mu}[f(U)]-kM^{2}\right) \nonumber \\
    &\leq 2\exp\left(-\frac{k^{2} M^{5}}{4L^{2}_{O}}\right)\leq 2\exp\left(-\frac{k^{2} M^{3}}{256f^{2}_{+}(\bar{N})}\right). \label{C19}
\end{align}
Here, we have considered $f(U)$ as $H_{LO}(U)$, used the fact that $\mathbb{E}[f(U)]\geq 2\pi\bar{n}M^2$, and set $\epsilon$ as $k M^2$, where $0<k<2\pi\bar{n}$ is some constant. 
The last inequality holds because the upper bound of $L_{O}$ is $8Mf_{+}(N)$. 
Adopting big $\Theta$ notation and rewrite Eq. \eqref{C19}, we complete the proof of Theorem 1 : 
\begin{align}
    &\Prob_{U\sim\mu}\left(H_{LO} \neq \Theta(M^{2}) \right) \leq \Prob_{U\sim\mu}\left(H_{LO}(U) \leq (2\pi\bar{n}-k)M^{2}\right) \leq \exp\left(-\Theta(M)\right),\\ &\Prob_{U\sim\mu}\left(H_{LO} = \Theta(M^{2}) \right) \geq 1 - \exp\left(-\Theta(M)\right).
\end{align}
\section{Proof of Theorem 2} \label{Appen D}
\subsection{Upper bound of optimal QFI without local operation}
Consider the input state that maximizes the QFI for a given $U$ among all possible states with a mean photon number $\bar{N}$. 
Here, we consider the case when there is no local operation which means $\boldsymbol{\phi}=\boldsymbol{0}$. We denote the corresponding QFI as $\mathcal{H}(U)$. By Eq. \eqref{A15}, $\mathcal{H}(U)$ is written as
\begin{align}
    \mathcal{H}(U) = \max_{\bar{\bm{n}}} H(U,0,\bar{\bm{n}}). \label{D1}
\end{align}
In Eq. \eqref{A18}, we mentioned that the upper bound of the QFI is $2M+4Mf_{+}(p_{\nu}\bar{N})$ where $p_{\nu}\equiv \max_{1\leq i\leq M}\abs{\sum_{a=1}^{M} U_{ai}}^{2}/M$, which depends on $U$. For our proof of Theorem 2, let us define a function $\mathcal{G}_{i}\equiv4M+8\bar{N}\abs{\sum_{a=1}^{M}U_{ai}}^{2}$. Using the above functions, the following inequalities can be derived:
\begin{align}
  &\mathcal{H}(U) \leq 2M + 4Mf_{+}(p_{\nu}\bar{N}) =2M+4M\left(p_{\nu}\bar{N} + \sqrt{p^{2}_{\nu}\bar{N}^{2} + p_{\nu}\bar{N}}\right) \leq 2M +2M\left(4p_{\nu}\bar{N} + 1\right)=\max_{i} \mathcal{G}_{i}(U), \label{D2}
\end{align}
which are the key ingredients in the proof of Theorem 2.

We remark that in Fig. 2 of the main text we consider QFI $H_{MO}(U)$ using a squeezed vacuum state into an optimized mode without local optimization, i.e., $\bm{\phi}=0$. 
Using Eq. \eqref{A18}, $H_{MO}(U)$ is given by
\begin{align} \label{D3}
H_{MO}(U)\equiv \max_{1\leq b\leq M} \left(2M + 4\abs{\sum_{a=1}^{M}U_{ab}}^{2}f_{+}(\bar{N})\right).
\end{align}
For large $M$ limit, one can easily show that the leading orders of $H_{MO}(U)$ and $\max_{i} \mathcal{G}_i (U)$ are equal. 
Due to the sandwich theorem, one can find that in the limit of large $M$, $H_{MO}(U)$ and $\mathcal{H}(U)$ have the same behavior.

\subsection{Expectation value and Lipschitz constant of $\mathcal{G}_1 (U)$ }
In this subsection we find the expectation value of $ \mathcal{G}_{1}(U)$ over Haar-random unitary matrix $U$:
\begin{align}
    \E_{U \sim \mu}\left[\mathcal{G}_{1}(U)\right] =4M + 8\bar{N}\E_{U \sim \mu}\left[\abs{\sum_{a=1}^{M} U_{a1}}^{2}\right]= 4M + 8\bar{N}\E_{U \sim \mu}\left[\sum_{a,b=1}^{M}U_{a1}{U^{*} _{b1}}\right]
\end{align}
Note that the expectation value over Haar-random only cares the term $\sum_{a,b=1}^{M}{U_{a1}}{U^{*}_{b1}}$. As a similar procedure in Sec. \ref{Appen C}, we can find the expectation value of $\sum_{a,b=1}^{M}{U_{a1}}{U^{*}_{b1}}$ over Haar-random unitary matrix $U$:
\begin{align} 
  \E_{U \sim \mu}\left[\sum_{a,b=1}^{M}U_{a1}{U^{*} _{b1}}\right] &= \left(\prod_{i=1}^{M} \int_{0}^{\infty} 2R_i dR_i e^{-{R_i}^2}\int_{0}^{2\pi} \frac{d\theta_i}{2\pi}\right)\left(\sum_{a,b=1}^{M}\frac{R_a R_b e^{i\theta_a} e^{-i\theta_b}}{\sum_{k=1}^{M}{R_k}^2}\right) \label{D4} \\
  &
  =  \left(\prod_{i=1}^{M} \int_{0}^{\infty} 2R_i dR_i e^{-{R_i}^2}\right)\left(\frac{\sum_{a=1}^{M} {R_a}^2}{\sum_{k=1}^{M}{R_k}^2}\right) = 1. 
\end{align}
Therefore, the expectation value of $ \mathcal{G}_{1}(U)$ over Haar-random unitary matrix $U$ is
\begin{align}
     \E_{U \sim \mu}\left[\mathcal{G}_{1}(U)\right] = 4M + 8\bar{N} = (4+8\bar{n})M.
\end{align}
Additionally, following the same procedure as the one in Sec. \ref{Appen C}, one can easily find that the upper bound of Lipschitz constant $L_{G1}$ of function $\mathcal{G}_{1}(U)$ is $ 16\bar{n}M^{2}$. Note that since every entry of the Haar-random unitary matrix $U$ has the same probability distribution \cite{meckes2019random}, $\E_{U \sim \mu}\left[\mathcal{G}_{1}(U)\right]= \E_{U \sim \mu}\left[\mathcal{G}_{2}(U)\right]=\cdots = \E_{U \sim \mu}\left[\mathcal{G}_{M}(U)\right]$ and $L_{G1}= L_{G2}=\cdots = L_{GM}$.

\subsection{Proof of Theorem 2}
In Theorem 2, we claim that if we do not apply local operations, most of the QFI cannot attain Heisenberg scaling. Therefore, we need to show that the probability that $\mathcal{H}(U)$ attains $\Theta(M^{2})$ is exponentially small. 
Instead of directly showing that that, we take a detour.
First, we use the concentration meausure inequality in Eq. \eqref{C18} with respect to $\mathcal{G}_{1}(U)$ :
\begin{align}
    \Prob_{U\sim\mu}\left(\abs{\mathcal{G}_{1}(U) - (4+8\bar{n})M} \geq k M^{2-\delta}\right)  \leq 2\exp\left(-\frac{ k^{2}M^{5-2\delta}}{4L^{2}_{G1}}\right) \leq 2\exp\left(-\frac{k^{2} M^{1-2\delta}}{1024\bar{n}^{2}}\right).
\end{align}
We set $\epsilon$ as $k M^{2-\delta}$ where $k$ and $\delta$ are constant $k>0$ and $0<\delta<1/2$. 
Second, using $\mathcal{H}(U) \leq \max_{i} \mathcal{G}_{i}(U)$, we set the inequalities among some probabilities. Let us denote the event $\left(\abs{\mathcal{G}_{i}(U) - (4+8\bar{n})M} \geq k M^{2}\right)$ as $Q_i$ and $\left(\abs{\mathcal{G}_{\nu}(U) - (4+8\bar{n})M} \geq k M^{2}\right)$ as $Q_{\nu}$. The relation between $Q_{\nu}$ and others is $Q_{\nu} \subset \bigcup_{i=1}^{M}Q_i$. Note that $\mathcal{G}_{\nu}(U)$ is one of the $\{\mathcal{G}_{i}(U)\}_{i=1}^{M}$. Therefore the following inequalities can be established :
\begin{align}
    &\Prob_{U\sim\mu}\left(\abs{\mathcal{H}(U) - (4+8\bar{n})M} \geq k M^{2-\delta}\right)\leq \Prob_{U\sim\mu}\left(\abs{\mathcal{G}_{\nu}(U) - (4+8\bar{n})M} \geq k M^{2-\delta}\right)   \leq \Prob_{U\sim\mu}\left(\bigcup_{i=1}^{M}Q_i\right) \label{D9} \\
    &\leq \sum_{i=1}^{M}\Prob_{U\sim\mu}\left(Q_i\right)= M\Prob_{U\sim\mu}\left(\abs{\mathcal{G}_{1}(U) - (4+8\bar{n})M} \geq k M^{2}\right) \leq \exp\left(-\frac{k^{2} M^{1-2\delta}}{1024\bar{n}^{2}}+ \ln 2M\right). \label{D10} 
\end{align}
The first inequality comes from the fact that $\mathcal{H}(U) \leq \mathcal{G}_{\nu}(U)$ and the equality comes from the fact that $\Prob_{U\sim\mu}\left(Q_i\right)$ is symmetric over index $i$. Note that the expectation values and Lipschitz constants of $\left\{\mathcal{G}_{i}(U)\right\}_{i=1}^{M}$ are all same. Finally, using the above inequalities, we complete the proof of Theorem 2:
\begin{align}
     &\Prob_{U\sim\mu}\left(\mathcal{H}(U) = \Theta(M^{2}) \right) \leq \Prob_{U\sim\mu}\left(\abs{\mathcal{H}(U) - (4+8\bar{n})M} \geq k M^{2-\delta}\right) \leq \exp\left[-\Theta(M)\right],\\
     &\Prob_{U\sim\mu}\left(\mathcal{H}(U) \neq \Theta(M^{2}) \right)  \geq 1 - \exp\left[-\Theta(M)\right].
\end{align}

\section{Optimality of homodyne measurement} \label{Appen E}
When estimating a single parameter $\theta$, the error of estimation $\Delta^{2} \theta$ is bounded by the classical Cram\'er-Rao lower bound as $\Delta^{2} \theta \geq 1/F$, where $F$ is the Fisher information defined as $F(\theta)=\sum_x[\partial P(x|\theta)/\partial \theta]^2/P(x|\theta)$, where $P(x|\theta)$ is the conditional probability of obtaining an outcome $x$ when the unknown parameter is $\theta$ \cite{kay1993fundamentals}. 
When $P\left(\vec{x} \vert \theta\right)$ follows a multivariate normal distribution with its $M$-dimensional first moment vector $\vec{\mu}_\text{cl}$ and $M\times M$ covariance matrix $\Sigma_\text{cl}$, the corresponding Fisher information is written as \cite{porat1986computation, kay1993fundamentals}
\begin{align}
    F = \pdv{\vec{\mu}_\text{cl}^\text{T}}{\theta}\Sigma_\text{cl}^{-1}\pdv{\vec{\mu}_\text{cl}}{\theta}, \label{E1}
\end{align}
where we have assumed that $\partial \Sigma_\text{cl}/\partial\theta=0$.

Meanwhile, for an $M$-mode Gaussian state, characterized by its $2M$-dimensional first moment vector $\vec{\mu}_\text{Q}$ and $2M\times 2M$ covariance matrix $\Sigma_\text{Q}$,
its QFI for a parameter $\theta$ is also written as \cite{oh2019optimal}
\begin{align}
    H = \pdv{\vec{\mu}_\text{Q}^\text{T}}{\theta}\Sigma_\text{Q}^{-1}\pdv{\vec{\mu}_\text{Q}}{\theta}, \label{E2}
\end{align}
where $\mu_i\equiv \text{Tr}[\hat{\rho}\hat{Q}_i]$ and $\Sigma_{ij}\equiv\text{Tr}[\hat{\rho}\{\hat{Q}_i-\mu_i,\hat{Q}_j-\mu_j\}]/2$ with the quadrature operator vector $\hat{Q}\equiv(\hat{x}_1,\dots,\hat{x}_M,\hat{p}_1,\dots,\hat{p}_M)^{\intercal}$.
Here, we again assumed that $\partial \Sigma_Q/\partial\theta=0$.

When we perform homodyne detection on a Gaussian state, the output probability distribution follows a normal distribution.
If we perform homodyne detection along $x$-axis for each mode, its output distribution follows an $M$-dimensional multivariate normal distribution with its first moment vector $(\mu_\text{HD})_i=(\mu_\text{Q})_{i}$ and $(\Sigma_\text{HD})_{ij}=(\Sigma_\text{Q})_{ij}$ for $1\leq i, j \leq M$.
Since the output probability distribution is a normal distribution, we can apply Eq. \eqref{E1} for our scheme. 
In Theorem 1, we consider a single-mode squeezed vacuum state as an input state which is squeezed along the $x$ axis with the mean photon number $\bar{N}$.
The state is injected into the first mode and other modes are in vacuum. Here we denote the state as $\ket{\Psi}=\ket{sqz(0, \sinh^2 r =\bar{N})}\otimes \ket{0} \cdots \ket{0}$. The state undergoes beam splitter array, phase shift operation and displacement encoding in regular sequence. The state right before the measurement is $\hat{D}(x)\hat{R}(\bm{\phi})\hat{U}\ket{\Psi}$. 
One can easily check that when the optimal phase shifts are applied, quantum covariance matrix's $M\times M$ off-diagonal block matrix vanishes, i.e., if we write
\begin{align}
    \Sigma_\text{Q}=
    \begin{pmatrix}
        \Sigma^{xx}_\text{Q} & \Sigma^{xp}_\text{Q} \\ 
        \Sigma^{px}_\text{Q} & \Sigma^{pp}_\text{Q}
    \end{pmatrix},
\end{align}
$\Sigma^{xp}_\text{Q}=\Sigma^{px}_\text{Q}=0$.
Noting that for our case, $\partial \vec{\mu}_\text{Q}/\partial x=(1,\dots,1,0,\dots,0)$, we can rewrite the QFI as
\begin{align}
    H=\sum_{i,j=1}^M[(\Sigma_\text{Q}^{xx})^{-1}]_{ij}.
\end{align}
Also, for homodyne detection, the covariance matrix becomes $\Sigma_\text{HD}=\Sigma^{xx}_\text{Q}$ we can find that
\begin{align}
    F=\sum_{i,j=1}^M[(\Sigma^{xx}_\text{Q})^{-1}]_{ij}.
\end{align}
Therefore, the classical Fisher information of homodyne detection is the same as the QFI, which shows that homodyne detection is optimal.

\section{Effect of photon loss} \label{Appen F}
Photon loss can be modeled by a beam splitter with its transmittivity $\eta$.
The beam splitter transforms annihilation operator as~$\hat{a}_j \rightarrow \sqrt{\eta} \hat{a}_j + \sqrt{1- \eta} \hat{e}_j$, where $\hat{e}_j$ is an annihilation operator for environment mode for all $j$'s \cite{PhysRevA.48.3265}.
When there is photon loss, the covariance matrix of Gaussian state transforms like \cite{serafini2017quantum} 
\begin{gather}
   \Sigma_\text{Q} \to \eta \Sigma_\text{Q} + (1-\eta)\frac{\mathbb{1}_{2M}}{2},
\end{gather}
where $\sqrt{\eta}$ is the transmittivity of the beam splitter. Using Eq.(\ref{E2}), the QFI becomes
\begin{align}
   H_{LO}(U,\eta=1)=2M + 4\left(\sum_{a=1}^{M}\abs{U_{a1}}\right)^{2}f_+(\bar{N})
   \to
   H_{LO}(U,\eta)=2M + 4\left(\sum_{a=1}^{M}\abs{U_{a1}}\right)^{2} \left[\frac{\eta f_{+}(\bar{N})}{2(1-\eta)f_{+}(\bar{N})+1}\right].
\end{align}
Thus, when there is photon loss, the change of QFI can be captured by
\begin{align}
    f_+(\bar{N})\to\frac{\eta f_{+}(\bar{N})}{2(1-\eta)f_{+}(\bar{N})+1}.
\end{align}
Let us find the tolerable $\eta$ still sustains $\E_{U \sim \mu}\left[H_{LO}(U,\eta)\right]=\Theta(M^{2})$.
We can find the bound using Taylor's theorem such that
\begin{align}
    \frac{\eta f_{+}(\bar{N})}{2(1-\eta)f_{+}(\bar{N})+1}\geq \eta f_+(\bar{N})-2(1-\eta)f_+(\bar{N})^2.
\end{align}
One can easily check that this can be larger than $\alpha f_+(\bar{N})$ with some constant $0<\alpha<1$ when

\begin{align}\label{eq:alpha}
    \eta\geq 1-\frac{1-\alpha}{1+2f_+(\bar{N})}.
\end{align}
In other words, when a loss rate $1-\eta$ satisfies
\begin{align}
    1-\eta \leq \beta, ~~~ \text{where} ~~~ \beta \equiv \frac{1-\alpha}{1+2f_+(\bar{N})}= \Theta\left(\frac{1}{\bar{n}M}\right),
\end{align}
we have
\begin{align}
    \frac{\eta f_{+}(\bar{N})}{2(1-\eta)f_{+}(\bar{N})+1}\geq \alpha f_+(\bar{N}),
\end{align}
or equivalently,
\begin{align}
   H_{LO}(U,\eta)=2M + 4\left(\sum_{a=1}^{M}\abs{U_{a1}}\right)^{2} \left[\frac{\eta f_{+}(\bar{N})}{2(1-\eta)f_{+}(\bar{N})+1}\right]
   \geq 2M + 4\alpha\left(\sum_{a=1}^{M}\abs{U_{a1}}\right)^{2}f_+(\bar{N}).
\end{align}
Therefore, the Heisenberg scaling is maintained. As $M$ grows, the loss rate decreases at least as $1/\bar{n}M$.

One can also directly prove the counterpart of Theorem 1.
Since the average over Haar-random matrix $U$ and Lipschitz constant rely only on the term $\abs{U_{a1}}\abs{U_{b1}}$, the expectation value of $H_{LO}(U,\eta)$ is 
\begin{align}
    \E_{U \sim \mu}\left[H_{LO}(U,\eta)\right] = 2M + 4\left(\frac{\pi}{4}(M-1)+1\right)\frac{\eta f_{+}(\bar{N})}{2(1-\eta)f_{+}(\bar{N})+1}
\end{align} and the corresponding Lipschitz constant $L_{O}(\eta)$ is 
\begin{align}
    L_{O}(\eta)= \frac{8M\eta f_{+}(\bar{N})}{2(1-\eta)f_{+}(\bar{N})+1}.
\end{align}
Thus, the concentration measure inequality Eq. \eqref{C19} becomes
\begin{align}
      \Prob_{U\sim\mu}\left(\abs{  H_{LO}(U,\eta) - \E_{U \sim \mu}[H_{LO}(U,\eta)]} \geq kM^{2}\right)  \leq 2\exp\left(-\frac{k^{2}M^{5}}{4L^{2}_{O}(\eta)}\right), \label{F3} 
\end{align}
where $k>0$ is a constant.
Note that the right-hand-side of Eq. \eqref{F3} is always exponentially small for any $0<\eta\leq 1$. 
Therefore, as far as $\E_{U \sim \mu}\left[H_{LO}(U,\eta)\right]$ is $\Theta(M^{2})$, Theorem 1 is still valid.
Hence, under photon-loss satisfying the condition of Eq.~\eqref{eq:alpha}, the Heisenberg scaling is maintained.

\section{Average distance between the covariance matrices after balanced BSN and random BSN}
In this section, we compare between the average distance to the optimal probe from the covariance matrices of random probes with and without local optimization.
The optimal probe state is generated by a single mode squeezed state and a balanced BSN. The local-optimized random probe state is generated by single mode squeezed state, random BSN, and local phase optimizations. Since all these probes are Gaussian state which have zero first moment vector, the quantum states can be characterized by their covariance matrix.
Therefore, we inspect the closeness of two states by using the Frobenius distance $\| A-B \|^{2}_{F}=\sum_{a,b}\abs{A_{ab}-B_{ab}}^{2}$ between the covariance matrices. 
Additionally, by comparing the optimal probe state and the random probe state without local optimization, we study the effect of the local phase optimizations on the closeness. 
We note that it would be more rigorous to consider the distance of quantum states such as the Bures distance or trace distance, which is however mathematically more involved, and that the direct relation between Frobenius norm of covariance matrices and the distance of quantum states is unclear, to the best of our knowledge.

%\cor{Through this section we inspect the closeness of the different kinds of the quantum probes by using the Frobenius distance between the covariance matrices. Besides the Frobenius norm, other distance measures like the Bures distance or trace-norm between the probes, also can be considered. We expect that such kinds of measures would show more direct connection between the different probes and their QFI, instead, the harder calculations might be required to find them.}

Re-ordering the quadrature operators as $\hat{R}=(\hat{x}_1,\hat{p}_1, \cdots, \hat{x}_M, \hat{p}_M)^{\intercal}$, yields the equivalent expression of the covariance matrix that we introduced in Sec. \ref{Appen E}, as follows:
\begin{align}
      \sigma_{ab}\equiv\text{Tr}[\hat{\rho}\{\hat{R}_a-R_a,\hat{R}_b-R_b\}]/2,
\end{align}
where $R_{a}\equiv\text{Tr}[\hat{\rho}\hat{R}_{a}]$.
Following the above notation, let us find the covariance matrices minus identity matrix of the probe states to find the distances between the matrices. Recall that we generate the probe states by injecting a single-mode squeezed vacuum state, whose squeezing parameter is $r$, into a global BSN represented by $M \times M$ random unitary matrix $U$ and operating local phase shift $\phi_a$ on $a$th mode. The covariance matrix minus identity is as follows (from now, for simplicity, we denote $\sigma - \mathbb{1}_{2M}$ as simply $\sigma$ and just call it covariance since subtracting the same matrix does not change the distance.):
\begin{align}
      &\sigma_{2a-1,2b-1}=  \Re\left[e^{i\phi_{a}}U_{a1}\right]\Re\left[e^{i\phi_{b}}U_{b1}\right]\left(\frac{e^{-2r}-1}{2}\right) +  \Im\left[e^{i\phi_{a}}U_{a1}\right]\Im\left[e^{i\phi_{b}}U_{b1}\right]\left(\frac{e^{2r}-1}{2}\right),\\
      &\sigma_{2a,2b}= \Im\left[e^{i\phi_{a}}U_{a1}\right]\Im\left[e^{i\phi_{b}}U_{b1}\right]\left(\frac{e^{-2r}-1}{2}\right) +  \Re\left[e^{i\phi_{a}}U_{a1}\right]\Re\left[e^{i\phi_{b}}U_{b1}\right]\left(\frac{e^{2r}-1}{2}\right),\\
      &\sigma_{2a-1,2b}=\Re\left[e^{i\phi_{a}}U_{a1}\right]\Im\left[e^{i\phi_{b}}U_{b1}\right]\left(\frac{e^{-2r}-1}{2}\right) +  \Im\left[e^{i\phi_{a}}U_{a1}\right]\Re\left[e^{i\phi_{b}}U_{b1}\right]\left(\frac{e^{2r}-1}{2}\right),\\
      &\sigma_{2a,2b-1}= \Im\left[e^{i\phi_{a}}U_{a1}\right]\Re\left[e^{i\phi_{b}}U_{b1}\right]\left(\frac{e^{-2r}-1}{2}\right) +  \Re\left[e^{i\phi_{a}}U_{a1}\right]\Im\left[e^{i\phi_{b}}U_{b1}\right]\left(\frac{e^{2r}-1}{2}\right).
\end{align}
Therefore, the covariance of a local-optimized random probe state by using the optimization condition, $e^{-i\phi^{*}_a}=U_{a1}/\abs{U_{a1}}$, is
\begin{align}
      \sigma^{LO}_{2a-1,2b-1}=  \left(\frac{e^{-2r}-1}{2}\right)\abs{U_{a1}}\abs{U_{b1}},~~~
      \sigma^{LO}_{2a,2b}=\left(\frac{e^{2r}-1}{2}\right)\abs{U_{a1}}\abs{U_{b1}},~~~
      \sigma^{LO}_{2a-1,2b}=0,~~~
      \sigma^{LO}_{2a,2b-1}=0,
\end{align}
and the covariance of the optimal probe state is
\begin{align}
      \sigma^{B}_{2a-1,2b-1}=\left(\frac{e^{-2r}-1}{2}\right)\frac{1}{M},~~~
      \sigma^{B}_{2a,2b}=\left(\frac{e^{2r}-1}{2}\right)\frac{1}{M},~~~
      \sigma^{B}_{2a-1,2b}=0,~~~
      \sigma^{B}_{2a,2b-1}=0.
\end{align}
Using the same Haar random calculation techniques that we used in Sec.~\ref{Appen C}, we find the expectation value over Haar random $U$ of the square of the Frobenius distance between $\sigma^{LO}$ and $\sigma^{B}$.
\begin{align}
    \E_{U \sim \mu}\left[\|\sigma^{LO}-\sigma^{B}\|^{2}_{F} \right]=2\cosh{2r}(\cosh{2r}-1)\left(1-\frac{1}{M}\left(1+\frac{\pi}{4}(M-1)\right)\right).
\end{align}
To investigate the effect of the local optimizations on the closeness, we get the covariance of a probe without local optimization, i.e., $\phi_a =0$ for all $a$:
\begin{align}
      &\sigma^{0}_{2a-1,2b-1}=  \Re\left[U_{a1}\right]\Re\left[U_{b1}\right]\left(\frac{e^{-2r}-1}{2}\right) +  \Im\left[U_{a1}\right]\Im\left[U_{b1}\right]\left(\frac{e^{2r}-1}{2}\right),\\
      &\sigma^{0}_{2a,2b}= \Im\left[U_{a1}\right]\Im\left[U_{b1}\right]\left(\frac{e^{-2r}-1}{2}\right) +  \Re\left[U_{a1}\right]\Re\left[U_{b1}\right]\left(\frac{e^{2r}-1}{2}\right),\\
      &\sigma^{0}_{2a-1,2b}=\Re\left[U_{a1}\right]\Im\left[U_{b1}\right]\left(\frac{e^{-2r}-1}{2}\right) +  \Im\left[U_{a1}\right]\Re\left[U_{b1}\right]\left(\frac{e^{2r}-1}{2}\right),\\
      &\sigma^{0}_{2a,2b-1}= \Im\left[U_{a1}\right]\Re\left[U_{b1}\right]\left(\frac{e^{-2r}-1}{2}\right) +  \Re\left[U_{a1}\right]\Im\left[U_{b1}\right]\left(\frac{e^{2r}-1}{2}\right).
\end{align}
Similarly, we get the average distance between covariances of random probes and that of balanced probe:
\begin{align}
    \E_{U \sim \mu}\left[\|\sigma^{0}-\sigma^{B}\|^{2}_{F} \right]=  2\cosh{2r}(\cosh{2r}-1) - \frac{\left(\cosh{2r}-1\right)^{2}}{M}.
\end{align}
Here, we have used the fact that $\sum_{a,b=1}^{M}\abs{U_{a1}}^{2}\abs{U_{b1}}^{2}=1$. We can compare the above expectation values with the trivial case. 
Consider the covariance of the trivial BSN, i.e., $U=\mathbb{1}_M$ which is denoted by $\sigma^{\mathbb{1}_M}$. The distance between the optimal probe and the trivial probe is then given by
\begin{align}
   \| \sigma^{\mathbb{1}_M}-\sigma^{B} \|^{2}_{F} = 2\cosh{2r}(\cosh{2r}-1)\left(1-\frac{1}{M}\right).
\end{align}
For fixed squeezing parameter $r$, as $M$ grows, $\E_{U \sim \mu}\left[\|\sigma^{LO}-\sigma^{B}\|^{2}_{F} \right]$, $\E_{U \sim \mu}\left[\|\sigma^{0}-\sigma^{B}\|^{2}_{F} \right]$ and $\| \sigma^{\mathbb{1}_M}-\sigma^{B} \|^{2}_{F}$ go to
\begin{align}
    \E_{U \sim \mu}\left[\|\sigma^{LO}-\sigma^{B}\|^{2}_{F} \right]&\to \left(2-\frac{\pi}{2}\right)\cosh{2r}(\cosh{2r}-1), \label{S88}\\ 
    \E_{U \sim \mu}\left[\|\sigma^{0}-\sigma^{B}\|^{2}_{F} \right]&\to 2\cosh{2r}(\cosh{2r}-1),\label{S89}\\
    \| \sigma^{\mathbb{1}_M}-\sigma^{B} \|^{2}_{F} &\to 2\cosh{2r}(\cosh{2r}-1).\label{S90}
\end{align}
Notably, the asymptotic average distance between local-optimized covariances and the optimal covariance is smaller than that between covariances without optimization and the optimal covariance.
Specifically, the ratio between the two averages $\E_{U \sim \mu}\left[\|\sigma^{LO}-\sigma^{B}\|^{2}_{F} \right]/\E_{U \sim \mu}\left[\|\sigma^{0}-\sigma^{B}\|^{2}_{F} \right]$ goes to $1-\pi/4<1$; thus, local optimization reduces the average distance between the covariance of random probes and the optimal probe.
In addition, the distance to the optimal one from random covariances without optimization is the same as the distance between the trivial case.
It shows that the local optimization indeed reduces the distance of the probe states to the optimal state.

\section{Two different descriptions for bosonic states}
\subsection{Notations for CV system and particles system}
In this section, we recall two different representations of a bosonic state which has definite photon number $N$ in $M$ modes, which we call mode formalism and particle formalism.
Let us consider a state that $i$th mode is occupied by $n_i$ photons, where $\sum_{i=1}^M n_i =N$.
For mode formalism, we denote such a quantum state as $\ket{n_1,\dots,n_M}$.
For the particle formalism, $\ket{i^p}_{j}$ means that $j$th particle is in the $i$th level of the system where $j \in \{0,\dots,N\}$ and $i \in \{1,\dots,M\}$. 
In addition, we define the permutation operator which symmetrizes a quantum state (with respect to particle exchange) to take account of the statistic of boson as $\text{P}[\ket{\cdots}]$. For example, 
\begin{align}
    \ket{2,1,0}=\text{P}\left[\ket{1^p}_{1}\ket{1^p}_{2}\ket{2^p}_{3}\right]=\frac{1}{\sqrt{3}}\left(\ket{1^p}_{1}\ket{1^p}_{2}\ket{2^p}_{3}+\ket{1^p}_{1}\ket{2^p}_{2}\ket{1^p}_{3}+\ket{2^p}_{1}\ket{1^p}_{2}\ket{1^p}_{3}\right).
\end{align}
Observe that the state has entanglement in particle formalism but does not have entanglement in mode formalism. 
More examples and basics concepts of mode and particle entanglement and their formalism can be found in Ref. \cite{Demkowicz-Dobrzanski2015}

\subsection{Distributed displacement sensing exploiting definite photon number state as an input state}
Let us revisit our distributed displacement sensing scheme with a single-mode definite number state $\ket{N,0,\dots,0}$ as an input state instead of a single-mode squeezed state. Using similar calculations used to derive \eqref{B4} (here, we adopt the same local-phase-optimization condition), one can show that the corresponding quantum Fisher information is
\begin{align} 
  H_{LO}(U,\bm{\phi}^*,\ket{N,0,\dots,0})=2M + 4N\left(\sum_{a=1}^{M}\abs{U_{a1}}\right)^{2}. 
\end{align}
Therefore, one can easily find that for single-mode number state, Lemma 1 and Theorem 1 are still valid simply changing $f_{+}(\bar{N})$ into $N$. This results implies that most of quantum probes $\ket{\psi_N}=\hat{R}(\bm{\phi}^*)\hat{U}\ket{N,0,\dots,0}$ are useful for distributed displacement sensing.

\subsection{Usefulness of $\ket{\psi_N}$ in terms of phase estimation}
From the paricle formalism, $\ket{N,0,\cdots,0}$ can be expressed as $\text{P}\left[\ket{1^p}_{1}\ket{1^p}_{2}\cdots \ket{1^p}_{N} \right]$ which is a product state for both mode and particle formalisms. 
However, the quantum probe $\ket{\psi_N}$ is a mode entangled state whereas it is a product state when the probe is described by particle formalism because the BSNs are local operations concerning for particle formalism \cite{Demkowicz-Dobrzanski2015}. We emphasize that local operations cannot change the entanglement of the system. Therefore the quantum probe $\ket{\psi_N}$ cannot attain Heisenberg scaling (in terms of particle number) in the phase estimation schemes like in Ref.~\cite{oszmaniec2016random}, even if any local unitary operations (operating on each particle) are allowed \cite{giovannetti2006quantum}.

%\cor{If we project a single-mode squeezed state on one of the number states, a similar discussion can be conducted on a single-mode squeezed state which is the input state in our main text \cite{Demkowicz-Dobrzanski2015}.}

\bibliography{reference.bib}